\documentclass[aps,reprint,superscriptaddress,longbibliography,nofootinbib]{revtex4-2}
\usepackage{graphicx}
\usepackage{dcolumn}
\usepackage{bm}
\usepackage{amsmath,amssymb,bm}
\usepackage{wasysym}
\usepackage{graphicx}
\usepackage{epstopdf}
\usepackage{latexsym}
\usepackage{subfigure}
\usepackage[usenames, dvipsnames]{color}
\usepackage[usenames, dvipsnames]{xcolor}
\usepackage{natbib}
\usepackage{braket}
\usepackage{float}
\usepackage[normalem]{ulem}
\usepackage{comment}
\usepackage{mathtools}
\usepackage{array}
\usepackage{tabu}
\usepackage{multirow}
\usepackage{chemformula}
\usepackage{svg}
\usepackage[T1]{fontenc}
\usepackage[colorlinks = true,
            linkcolor = red,
            urlcolor  = titleblue,
            citecolor = cgreen,
            anchorcolor = blue]{hyperref}
\newcommand\redsout{\bgroup\markoverwith{\textcolor{red}{\rule[0.5ex]{2pt}{0.4pt}}}\ULon}
\newcommand\bluesout{\bgroup\markoverwith{\textcolor{black}{\rule[0.5ex]{2pt}{0.4pt}}}\ULon}

\newcommand{\prlsec}[1]{{{\it #1}}}
\definecolor{titleblue}{RGB}{37,96,167}
\definecolor{cgreen}{RGB}{11,200,10}

\def\acz2{ac-$\mathbb{Z}_2$}

\begin{document}
\raggedbottom
\title{\textcolor{black}{Quasi-particle sector hybridization and the onset of chaos in spin ladders}\\
}

\author{Arnav Pushkar\,\,\href{https://orcid.org/0000-0002-1978-8307}
{\includegraphics[scale=0.05]{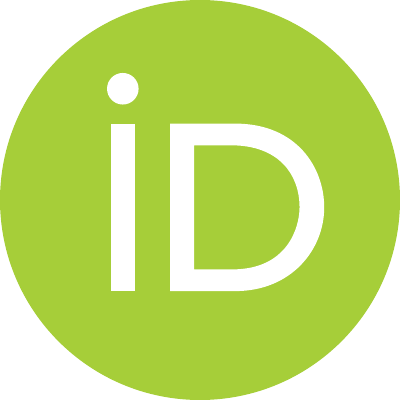}}}
\email{arnavpushkar@gmail.com}
\affiliation{Department of Physics, Indian Institute of Technology Bombay, Mumbai 400076, India.}
\affiliation{Department of Physics, Kyoto University, Kitashirakawa Oiwakecho, Sakyo-ku, Kyoto 606-8502, Japan}
\author{Abhik Kumar Saha\,\,\href{https://orcid.org/0000-0001-8168-6742}
{\includegraphics[scale=0.05]{orcidid.pdf}}}
\email{saha.abhikkumar.3k@kyoto-u.ac.jp}
\affiliation{Department of Physics, Kyoto University, Kitashirakawa Oiwakecho, Sakyo-ku, Kyoto 606-8502, Japan}

\begin{abstract}
How quasiparticle numbers influence the emergence of quantum chaos is a fundamental question in interacting many-body systems. 
We investigate this interplay in a spin-$1/2$ XXZ ladder, where the strong-rung regime supports an approximate organization of many-body eigenstates  into quasi-particle (triplon) number sectors. 
Using spectral statistics, triplon-based observables, and von Neumann entanglement entropy, we characterize the microscopic hybridization of triplon-number sectors as the leg coupling increases. 
We find that the crossover toward Gaussian orthogonal ensemble statistics shifts to weaker couplings with increasing system size, whereas the characteristic crossover scale associated with triplon-sector hybridization exhibits a much weaker size dependence. 
Furthermore, eigenstate-to-eigenstate fluctuations exhibit scaling consistent with the eigenstate thermalization hypothesis even while remnants of the strong-rung structure persist, suggesting that spectral chaos does not necessarily require complete sector hybridization.
\end{abstract}
\maketitle

\prlsec{Introduction.} 
Understanding how isolated quantum many-body systems develop quantum-chaotic and thermal behavior remains a fundamental problem in quantum statistical mechanics~\cite{NandkishoreHuse2015,GogolinEisert2016}. 
The eigenstate thermalization hypothesis (ETH) provides a framework for understanding thermalization through the properties of individual many-body eigenstates \cite{Deutsch1991,Srednicki,RigolNature,Deutsch2018,DAlessio2016,MierzejewskiVidmar2020}, while random-matrix theory provides statistical signatures of quantum chaos through energy-level correlations \cite{OganesyanHuse2007,Atas2013,BohigasGiannoniSchmit1984}. 
Although these approaches have significantly advanced our understanding of quantum many-body systems \cite{SantosRigol2010, PalHuse2010,LeBlondSelsPolkovnikovRigol2021, RichterDymarskySteinigewegGemmer2020}, an important question remains as to how the microscopic structure of eigenstates change as interactions drive a system toward chaotic behavior. 
In systems with identifiable local excitations, many-body eigenstates retain an approximate organization into quasi-particle number sectors~\cite{KnetterSchmidtUhrig2003, SchmiedinghoffEtAl2022}.
A recent study of the staggered-field XXZ chain has connected the weakening of quantum-chaotic level statistics with the emergence of increasingly well-defined approximate domain-wall-number sectors~\cite{WildeboerLajerKonik2026}.
An important question is whether quantum chaotic spectral correlations can develop while this approximate structure remains discernible.
In particular, interactions can induce hybridization between states characterized by approximately conserved quantum numbers, progressively modifying the organization of many-body eigenstates~\cite{BulchandaniHuseGopalakrishnan2022}. 
Related studies have shown that approximately conserved quasi-particle numbers can lead to long-lived prethermal regimes~\cite{BirnkammerBastianelloKnap2022,PawlowskiPanfilHerbrychMierzejewski2024,MallayyaRigolDeRoeck2019}.

\begin{figure}[htbp]
\centering
\includegraphics[width=0.95\linewidth]{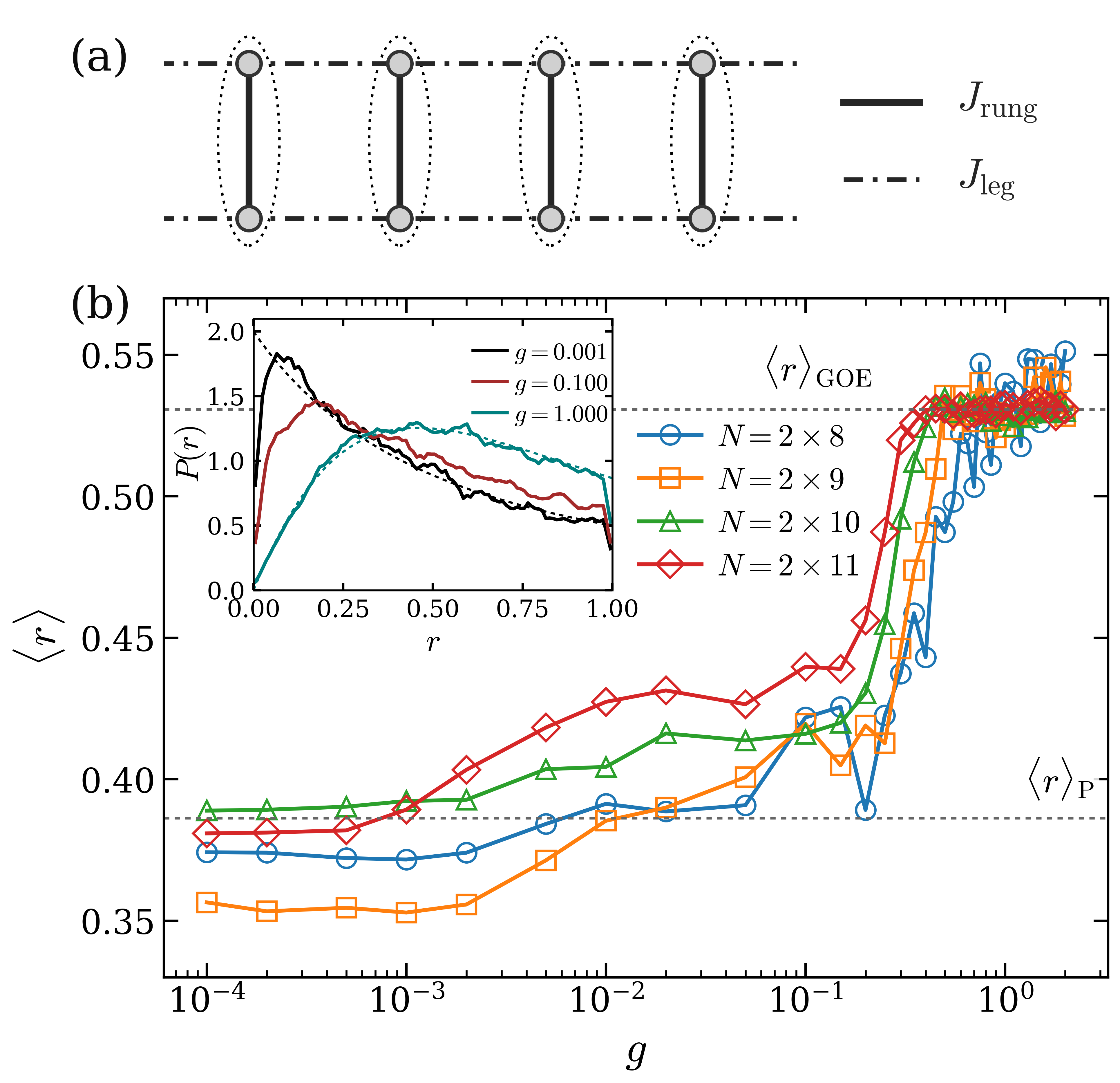}
\caption{(a) Schematic of the spin-$1/2$ XXZ ladder, with solid and dash-dotted lines representing the rung ($J_{\rm rung}$) and leg ($J_{\rm leg}$) couplings, respectively. 
(b) Average adjacent-gap ratio $\langle r\rangle$ as a function of $g$ for different system sizes at $\Delta=0.8$, calculated in the $S^z_{\rm tot}=0$, $k=0$ sector using the central spectral window $0.40\leq\mathcal{E}\leq0.60$. 
Horizontal dotted lines indicate the Poisson and GOE benchmarks. 
Inset: Adjacent-gap ratio distributions $P(r)$ for three representative coupling strengths at $N=2\times11$, with dashed curves indicating the analytical predictions for Poisson and GOE statistics.}
\label{fig:r_value}
\end{figure}

Quantum spin ladders \cite{BarnesEtAl1993, DagottoRice1996,PhysRevB.53.8521,PhysRevE.93.042101,85kv-972r} provide a natural setting for investigating the interplay between interactions, eigenstate structure, and quantum chaos \cite{McClarty2020,IadecolaZnidaric2019,NajafzadehLangari2026,PhysRevLett.110.070602,PhysRevB.88.205135}. 
Their properties depend on the relative strengths of the couplings along the legs and across the rungs (see Fig.~\ref{fig:r_value}(a)), allowing different interaction regimes to be explored within a single model. 
When the rung coupling dominates, the local singlet and triplet states provide a natural basis for describing many-body eigenstates~\cite{SachdevBhatt1990,GopalanRiceSigrist1994}. 
The rung-triplet excitations, called \emph{triplons}, provide a quasi-particle description in this limit~\cite{KnetterSchmidtGruningerUhrig2001, KnetterSchmidtUhrig2003}, with the eigenstates approximately organized into triplon-number sectors.
Previous studies of spin ladders have revealed nonthermal phenomena, including many-body localization \cite{BayganLimSheng2015, HuHongGuoRen2026, RichterPal2022}, anomalous eigenstates \cite{IadecolaZnidaric2019}, and ergodicity breaking \cite{NajafzadehLangari2026}. 
In fully frustrated ladders, exact conservation of individual rung spins gives rise to distinct sectors with different thermalization properties~\cite{McClarty2020}, while kinetically constrained Rydberg ladders can host quantum many-body scars and persistent imbalances that violate ETH~\cite{PalEtAl2025,scars_review}.
These studies illustrate how conservation laws, frustration, and kinetic constraints influence the eigenstate structure and ergodic behavior. 
However, a different situation arises when rung-based quantum numbers are only approximately conserved and interactions progressively hybridize the corresponding quasi-particle number sectors.
This raises a natural question: \textit{How does the breakdown of approximate rung-based conservation influence the microscopic structure of many-body eigenstates as quantum chaos emerges?} 
We address this question in a spin-$1/2$ XXZ ladder by varying the leg-to-rung coupling ratio. 

To characterize the hybridization between triplon-number sectors, we introduce triplon-based diagnostics involving triplon-number fluctuations and connected nearest-neighbor triplon correlations, complemented by level-spacing statistics and von Neumann entanglement entropy. 
Our results reveal that increasing the leg coupling enhances triplon-sector hybridization, accompanied by smoother energy-dependent eigenstate distributions and eigenstate-to-eigenstate fluctuations consistent with ETH scaling. 
Interestingly, the crossover toward Gaussian orthogonal ensemble (GOE) statistics shifts to weaker couplings with increasing system size, whereas the characteristic crossover scales associated with triplon-number fluctuations exhibit a much weaker size dependence. 
These findings suggest that GOE-like spectral correlations can develop while remnants of the approximate triplon-sector organization persist, providing a microscopic perspective on the emergence of quantum chaos in spin ladders.

\prlsec{Model.}  We consider a spin-$1/2$ $\rm XXZ$ ladder consisting of $N=2\times L$ spins arranged on $L$ rungs, with periodic boundary conditions along the leg direction (see Fig.~\ref{fig:r_value}(a)). 
The Hamiltonian is given by
\begin{equation}
\label{eqn:Ham}
\begin{aligned}
H={}& J_{\rm rung}\sum_{i=1}^{L}
h^{\rm XXZ}_{(i,1),(i,2)}\\
&+J_{\rm leg}\sum_{i=1}^{L}
\Big[
h^{\rm XXZ}_{(i,1),(i+1,1)}
+h^{\rm XXZ}_{(i,2),(i+1,2)}
\Big],
\end{aligned}
\end{equation}
where $h^{\rm XXZ}_{a,b}=\frac12(S_a^+S_b^-+S_a^-S_b^+)+\Delta S_a^zS_b^z$, with $J_{\rm rung}>0$ and $J_{\rm leg}\geq0$ denoting the rung and leg couplings, respectively. 
We fix the anisotropy at $\Delta=0.8$. 
The Hamiltonian conserves total magnetization $S^z_{\rm tot}=\sum_{i=1}^{L}(S^z_{i,1}+S^z_{i,2})$. 
We additionally include weak perturbations that break other discrete symmetries while preserving $S^z_{\rm tot}$ and translation by one rung~\cite{SantosRigolSymmetries2010}. 
These perturbations are incorporated throughout our numerical calculation, with their explicit forms and strengths provided in the Supplemental Material~\cite{supp_info}. 
We work in the $S^z_{\rm tot}=0$ and zero-momentum ($k=0$) sector, considering system sizes up to $N=2\times11$, for which the Hilbert space dimension of the selected symmetry sector is $64132$. 
We investigate the competition between rung and leg interactions by varying the dimensionless coupling ratio $g=J_{\rm leg}/J_{\rm rung}$. 
In the strong-rung limit ($g\ll1$), the eigenstates retain an approximate organization in terms of rung-singlet and triplet excitations \cite{rungnote}, whereas for $g\sim1$, the competing interactions lead to substantial hybridization of the rung-based states.

\prlsec{Level Statistics.} We first investigate how the competition between rung and leg couplings influences the spectral properties of the ladder. 
To characterize the emergence of quantum-chaotic level correlations, we consider the adjacent-gap ratio of consecutive energy levels~\cite{Atas2013,OganesyanHuse2007},
\begin{equation}
\label{eqn:r_value}
r_n=\frac{\min(\delta_n,\delta_{n+1})}{\max(\delta_n,\delta_{n+1})},
\qquad \delta_n=E_{n+1}-E_n,
\end{equation}
where $E_n$ denotes the $n$-th eigenvalue of the ordered many-body spectrum. 
The average gap ratio is defined as $\langle r\rangle=N_{\mathcal W}^{-1}\sum_{n\in\mathcal W}r_n$, where $\mathcal W$ denotes the set of adjacent-gap ratios evaluated within the chosen spectral window, and $N_{\mathcal W}$ is their total number. 
Throughout our calculations, we consider the central $20\%$ of the normalized spectral bandwidth, corresponding to $0.40\leq\mathcal{E}\leq0.60$, where $\mathcal{E}=(E-E_{\min})/(E_{\max}-E_{\min})$ is the normalized energy, $E_{\min}$ and $E_{\max}$ denoting the minimum and maximum eigenenergies within the selected symmetry sector, respectively. 
The average gap ratio distinguishes uncorrelated energy levels, characterized by the Poisson value $\langle r\rangle_{\rm P}\approx0.386$, from spectra exhibiting level repulsion, characterized by the GOE value $\langle r\rangle_{\rm GOE}\approx0.5307$ \cite{Atas2013}. 

Fig.~\ref{fig:r_value}(b) shows the average gap ratio $\langle r\rangle$ as a function of the coupling ratio $g$ for different system sizes. 
In the strong-rung regime ($g\ll1$), $\langle r\rangle$ remains close to the Poisson value, indicating relatively weak level repulsion. 
As the leg coupling increases, interactions between neighboring rungs increasingly hybridize the many-body states, leading to stronger level repulsion. 
Consequently, $\langle r\rangle$ approaches the GOE value, indicating the development of quantum-chaotic spectral correlations~\cite{RabsonNarozhnyMillis2004}. 
Notably, the crossover toward GOE statistics shifts to smaller values of $g$ with increasing system size. 
This trend suggests that progressively weaker leg couplings are sufficient to produce appreciable level repulsion as the many-body Hilbert space grows~\cite{SzaszSchagrinPozsgayTakacs2021,LeBlondSelsPolkovnikovRigol2021}.

The inset of Fig.~\ref{fig:r_value}(b) further illustrates the crossover from Poisson to GOE level statistics through the probability distribution $P(r)$ of adjacent-gap ratios at representative coupling strengths. 
At $g=0.001$, the distribution exhibits enhanced weight at small $r$, consistent with weak level repulsion and approximately Poisson-like statistics. 
For the intermediate coupling, $g=0.1$, the distribution gradually shifts away from the Poisson form, indicating the development of correlations between neighboring energy levels. 
At $g=1.0$, the suppression of small gap ratios and the agreement with the GOE distribution provide further evidence of strong level repulsion. 
Together, these results establish a finite-size crossover from approximately Poisson to GOE spectral statistics as the leg coupling increases. 
However, spectral statistics alone do not reveal how the microscopic organization of many-body eigenstates changes across this crossover. 
We therefore introduce triplon-based diagnostics to characterize the hybridization between different triplon-number sectors.

\prlsec{Triplon Diagnostics.} 
In the strong-rung regime, the many-body eigenstates retain an approximate organization according to the number of rung-triplet excitations. 
To characterize this structure, we introduce the local triplet projector on rung $i$,
\begin{equation}
P_i^t=\mathbf S_{i,1}\cdot\mathbf S_{i,2}+\frac34,
\end{equation}
where $\mathbf{S}_{i,\ell}=(S^x_{i,\ell},S^y_{i,\ell},S^z_{i,\ell})$
denotes the spin-$1/2$ operator on rung $i$ and leg $\ell=1,2$. 
The projector has eigenvalues $0$ and $1$ for rung-singlet and rung-triplet states ~\cite{rungnote}. The total triplon-number operator is then given by
\begin{equation}
N_t=\sum_i P_i^t.
\end{equation}

As the leg coupling increases, interactions between neighboring rungs hybridize different triplon-number sectors, making the total triplon number less well defined in individual eigenstates. 
We quantify this hybridization through the eigenstate triplon-number variance,
\begin{equation}
\Delta N_{t,\alpha}^2
=\langle N_t^2\rangle_\alpha
-\langle N_t\rangle_\alpha^2,
\end{equation}
where $\langle\mathcal{O}\rangle_\alpha
=\langle E_\alpha|\mathcal{O}|E_\alpha\rangle$
denotes the expectation value of an operator $\mathcal{O}$ in the many-body eigenstate $|E_\alpha\rangle$. 
The variance vanishes for states with definite triplon number, whereas a nonzero value indicates hybridization from different triplon-number sectors. 
For reference, statistically independent rung occupations yield the thermodynamic infinite-temperature value $\Delta N_{t,\alpha}^2/L=3/16$.

To further characterize the spatial organization of triplet occupations, we introduce the connected nearest-neighbor triplon correlation,
\begin{equation}
\label{eq:Cconn}
\begin{aligned}
C_{tt}^{(1),{\rm conn}}
=\frac{1}{L}\sum_i\big[
\langle P_i^tP_{i+1}^t\rangle_\alpha
-\langle P_i^t\rangle_\alpha
\langle P_{i+1}^t\rangle_\alpha
\big].
\end{aligned}
\end{equation}
While the triplon-number variance captures fluctuations in the total triplet occupation, including correlations at all rung separations, $C_{tt}^{(1),{\rm conn}}$ specifically probes correlation between neighboring rung occupations, providing complementary information about their spatial organization~\cite{note_triplet_occupation}.

\begin{figure}[htbp]
\centering
\includegraphics[width=\linewidth]{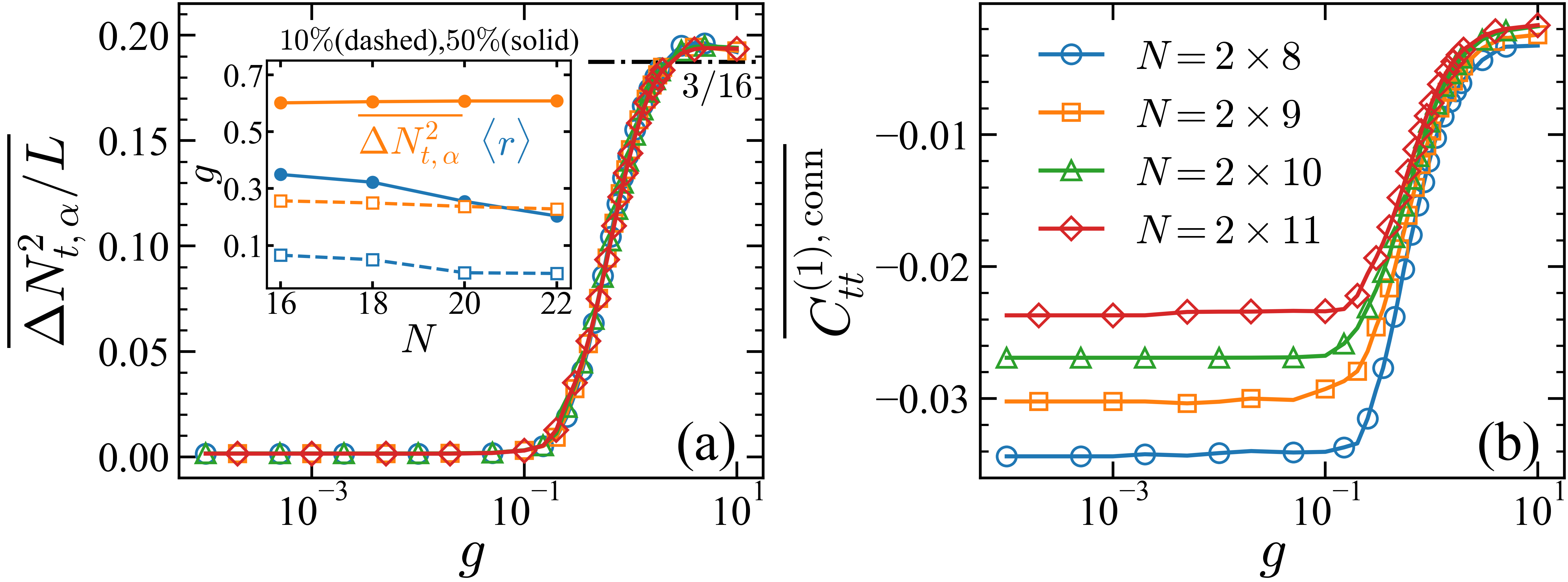}
\caption{Triplon-based diagnostics as a function of $g$ for different system sizes at $\Delta=0.8$. 
(a) Eigenstate-averaged triplon-number variance per rung, $\overline{\Delta N_{t,\alpha}^2/L}$. The horizontal line denotes the infinite-temperature benchmark $3/16$. 
(b) Eigenstate-averaged connected nearest-neighbor triplon correlation, $\overline{C_{tt,\alpha}^{(1),{\rm conn}}}$. 
The averages are taken over eigenstates within the central spectral window $0.40\leq\mathcal{E}\leq0.60$ in the $S^z_{\rm tot}=0$, $k=0$ sector. 
Inset of (a): System-size dependence of the characteristic coupling scales $g_{\mathcal O}^{10}$ (dashed) and $g_{\mathcal O}^{50}$ (solid), corresponding to $10\%$ and $50\%$ of the crossover in $\mathcal O=\langle r\rangle$ and $\overline{\Delta N_{t,\alpha}^2}$, respectively.}
\label{fig:triplon_diagnostics}
\end{figure}

Fig.~\ref{fig:triplon_diagnostics}(a) shows the eigenstate-averaged triplon-number variance, $\overline{\Delta N_{t,\alpha}^2/L}$, as a function of $g$, where the overbar denotes an average over eigenstates within the selected spectral window. 
In the strong-rung regime, the variance remains close to zero, reflecting an approximately definite triplon number. 
As $g$ increases, it grows rapidly and approaches the infinite-temperature benchmark $3/16$, indicating increasing hybridization between different triplon-number sectors. 
The curves exhibit relatively weak system-size dependence. 
To compare this behavior with the spectral crossover, the inset of Fig.~\ref{fig:triplon_diagnostics}(a) shows the characteristic coupling scales $g_{\mathcal O}^{10}$ and $g_{\mathcal O}^{50}$ for $\mathcal O=\langle r\rangle$ and $\overline{\Delta N_{t,\alpha}^2}$. 
These scales correspond to the couplings at which each observable reaches $10\%$ and $50\%$ of its total change between the weak- and strong-coupling reference values, respectively. 
Interestingly, the crossover scales extracted from $\langle r\rangle$ shift toward smaller $g$ with increasing system size, whereas those associated with $\Delta N_{t,\alpha}^2$ remain nearly unchanged. 
This suggests that GOE spectral correlations can develop before triplon-sector hybridization is complete, leaving remnants of the approximate rung-based structure. 
Fig.~\ref{fig:triplon_diagnostics}(b) represents the averaged connected nearest-neighbor triplon correlation, $\overline{C_{tt}^{(1),{\rm conn}}}$, as a function of $g$. 
In the strong-rung regime, the correlation is negative, indicating anticorrelations between neighboring triplet occupations, whose magnitude decreases with increasing system size. 
As $g$ increases, the correlation approaches zero, reflecting a progressive weakening of these anticorrelations. 
Together, these results show that triplon-sector hybridization is accompanied by changes in the spatial correlations of triplet occupations. 
To examine this hybridization at the level of individual many-body eigenstates, we next consider the corresponding eigenstate-resolved distributions.

\begin{figure}[htbp]
\centering
\includegraphics[width=\linewidth]{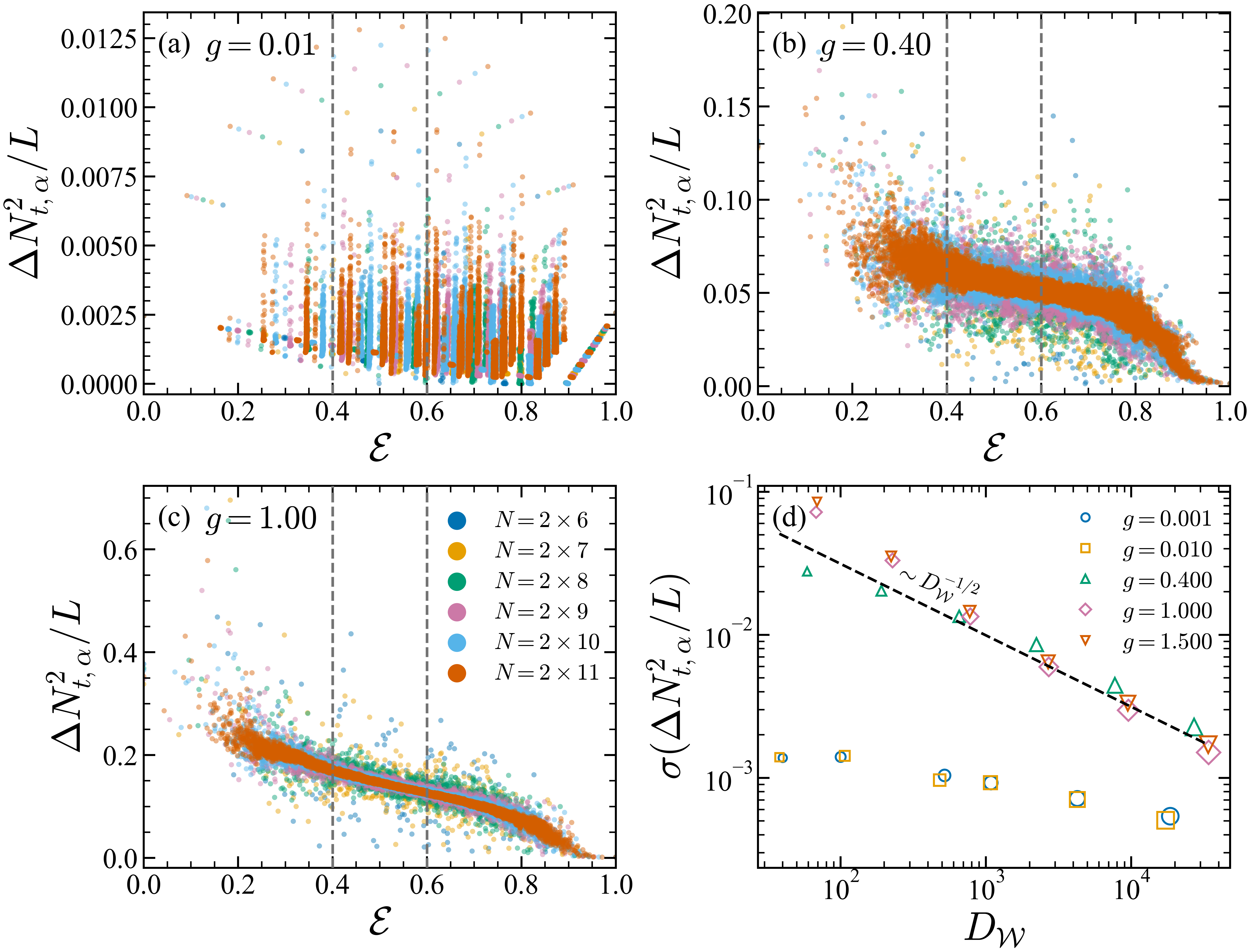}
\caption{Eigenstate-resolved triplon-number variance per rung, $\Delta N_{t,\alpha}^2/L$, as a function of $\mathcal{E}$ for different system sizes at (a) $g=0.01$, (b) $g=0.4$, and (c) $g=1.0$. 
Vertical dashed lines indicate the central spectral window $0.40\leq\mathcal{E}\leq0.60$. 
(d) Standard deviation $\sigma(\Delta N_{t,\alpha}^2/L)$ of the residual eigenstate-to-eigenstate fluctuations after subtracting a smooth fit to the energy dependence, plotted against the number of eigenstates $D_{\mathcal W}$ within the selected spectral window for different coupling strengths. 
The black dashed line indicates the ETH scaling $\sim D_{\mathcal W}^{-1/2}$.}
\label{fig:eev_dNt2}
\end{figure}

Figs.~\ref{fig:eev_dNt2}(a)-(c) show $\Delta N_{t,\alpha}^2/L$ as a function of $\mathcal{E}$ for three representative coupling strengths. 
In the strong-rung regime ($g=0.01$), the eigenstates exhibit distinct branches with small triplon-number fluctuations, reflecting their approximate organization into triplon-number sectors. 
At intermediate coupling ($g=0.4$), these branches become less distinguishable, giving rise to a broader, more continuous energy-dependent distribution. 
For $g=1.0$, the distribution becomes smoother, indicating substantial hybridization between different triplon-number sectors. 
A similar behavior is observed in the eigenstate-resolved connected triplon correlations, presented in the Supplemental Material~\cite{supp_info}. 
To quantify the eigenstate-to-eigenstate fluctuations, we subtract the smooth energy dependence of $\Delta N_{t,\alpha}^2/L$ within the central spectral window and calculate the standard deviation of the residuals, $\sigma(\Delta N_t^2/L)$. 
Fig.~\ref{fig:eev_dNt2}(d) shows its dependence on the number of eigenstates $D_{\mathcal W}$ within this window. 
For $g=1.0$ and $g=1.5$, the fluctuations decrease consistently with the ETH scaling $\sim D_{\mathcal W}^{-1/2}$~\cite{Beugeling_etal_pre_2014,Srednicki1999}.
The results at $g=0.4$ also approximately follow this scaling, indicating the development of ETH-consistent eigenstate fluctuations at intermediate coupling. 
In contrast, the weak-coupling results ($g=0.001$ and $g=0.01$) exhibit a much weaker system-size dependence and remain clearly separated from the ETH scaling behavior. 
These observations suggest that eigenstate fluctuations can exhibit ETH scaling even while remnants of the approximate triplon-sector organization persist.

\begin{figure}[htbp]
    \centering
    \includegraphics[width=\linewidth]{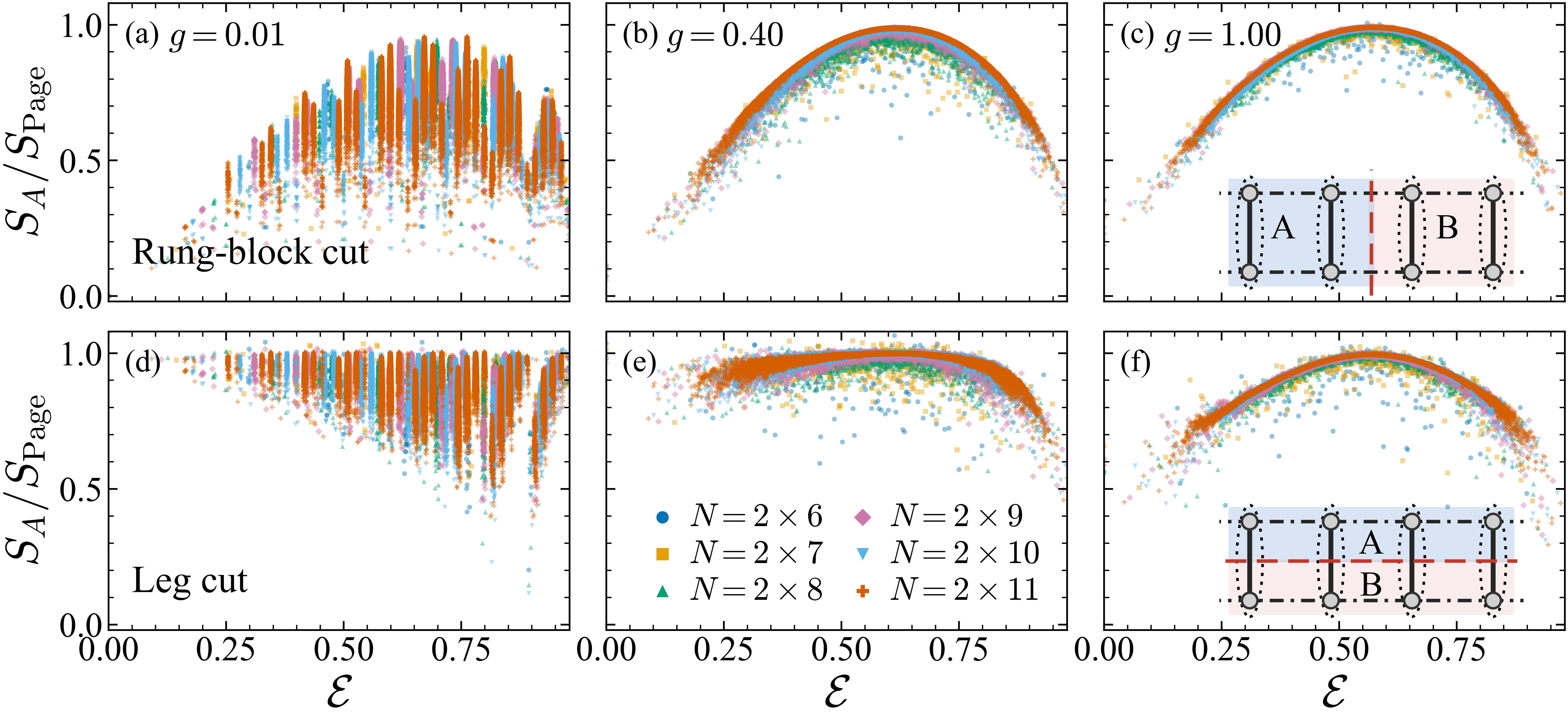}
    \caption{Eigenstate-resolved von Neumann entanglement entropy $S_A$, normalized by the corresponding Page value $S_{\rm Page}$, as a function of the normalized energy $\mathcal{E}$ for different system sizes. 
    The upper panels (a)–(c) show the rung-block cut, while the lower panels (d)–(f) show the leg cut, for $g=0.01$, $0.4$, and $1.0$, respectively. 
    The insets in (c) and (f) schematically illustrate the two bipartitions: consecutive complete rungs for the rung-block cut and one entire leg for the leg cut. 
    Each point represents an individual many-body eigenstate, with different colors and markers denoting system sizes.}
    \label{fig:ee_eigenstate_entanglement}
\end{figure}

\prlsec{Entanglement Entropy.} To further characterize the microscopic hybridization of many-body eigenstates across the crossover, we study the bipartite von Neumann entanglement entropy $S_A=-\mathrm{Tr}(\rho_A\ln\rho_A)$, where $\rho_A=\mathrm{Tr}_{B}(|E_\alpha\rangle\langle E_\alpha|)$ is the reduced density matrix of subsystem $A$ and $\mathrm{Tr}_{B}(\bullet)$ denotes the partial trace over subsystem $B$. 
We consider two distinct bipartitions of the ladder: the rung-block cut and the leg cut, as illustrated schematically in the insets of Figs.~\ref{fig:ee_eigenstate_entanglement}(c) and (f). In the rung-block cut, subsystem $A$ consists of consecutive complete rungs, probing entanglement between different groups of rungs. 
In contrast, the leg cut divides the ladder into its two legs, with subsystem $A$ comprising one entire leg. 
Since this partition separates the two spins on every rung, it is sensitive to the entanglement associated with the local rung-singlet and triplet structure. 
These two cuts therefore allow us to distinguish entanglement between different rungs from that already present within individual rungs. 
Fig.~\ref{fig:ee_eigenstate_entanglement} shows $S_A/S_{\rm Page}$ as a function of $\mathcal{E}$ for the two bipartitions, where $S_{\rm Page}$ denotes the average entanglement entropy of Haar-random pure states for the corresponding bipartition~\cite{Page1993}. $S_{\rm Page}$ is evaluated numerically by averaging the entanglement entropy over 50 normalized random real states within the $S^z_{\rm tot}=0$, $k=0$ symmetry sector, separately for each system size and bipartition.

For the rung-block cut (Figs.~\ref{fig:ee_eigenstate_entanglement}(a)-(c)), the entanglement exhibits distinct eigenstate branches in the strong-rung regime ($g=0.01$), consistent with the approximate organization into triplon-number sectors. 
Although some eigenstates already exhibit substantial entanglement, their distribution remains strongly structured. 
As $g$ increases, these branches progressively hybridize into a smoother energy-dependent distribution, with the entanglement approaching the Page benchmark near the middle of the spectrum (Figs.~\ref{fig:ee_eigenstate_entanglement}(b,c)). 
This behavior parallels the triplon-sector hybridization observed in the eigenstate-resolved fluctuations and is consistent with the entanglement properties of quantum-chaotic eigenstates~\cite{VidmarRigol2017,HaqueMcClartyKhaymovich2022}.

In contrast, the leg-cut entanglement is already substantial in the strong-rung regime (Fig.~\ref{fig:ee_eigenstate_entanglement}(d)), even when hybridization between different triplon-number sectors remains weak. 
This originates from the entanglement between the two spins forming each rung, which contributes to the leg-cut entropy even when correlations between different rungs are weak. 
Thus, a large leg-cut entropy alone does not necessarily indicate quantum-chaotic behavior. 
As $g$ increases, the eigenstate-resolved entropies develop a smoother energy-dependent distribution, approaching the Page benchmark near the center of the spectrum (Figs.~\ref{fig:ee_eigenstate_entanglement}(e,f)). 
Together, these results show that triplon-sector hybridization is accompanied by changes in the eigenstate-resolved entanglement structure.

\prlsec{Outlook and discussion.} 
We have investigated how the breakdown of approximate triplon-number conservation influences the microscopic structure of many-body eigenstates as quantum chaos emerges in a spin-$1/2$ XXZ ladder. 
In the strong-rung regime, eigenstates retain an approximate organization into triplon-number sectors, which progressively hybridize as the leg coupling increases. 
This evolution is reflected in the growth of triplon-number fluctuations, the development of smoother energy-dependent eigenstate distributions, and changes in bipartite entanglement across different cuts. 
Remarkably, the emergence of GOE spectral statistics does not necessarily coincide with the complete triplon-number sectors hybridization. 
While the spectral crossover shifts toward smaller couplings with increasing system size, the characteristic crossover scales associated with triplon-number fluctuations exhibit a much weaker system-size dependence. 
Furthermore, eigenstate-to-eigenstate fluctuations at intermediate coupling already display scaling consistent with ETH, even while remnants of the strong-rung structure persist. 
These findings suggest that quantum-chaotic behavior can emerge while approximate triplon-sector hybridization remains discernible, highlighting the importance of examining microscopic eigenstate structure alongside conventional spectral diagnostics.

Our findings motivate further investigations into the interplay between quasi-particle structure and quantum chaos~\cite{RuidasRoyBhattacharjeeMoessner2026, WildeboerLajerKonik2026,BilitewskiBhattacharjeeMoessner2021,KimHuse2013}. 
An important question is whether the different crossover scales persist toward the thermodynamic limit \cite{PhysRevB.105.L220405}.
Another promising direction is to investigate the non-equilibrium dynamics of initial states with well-defined quasiparticle number, particularly the redistribution of weight among different quasiparticle number sectors and the accompanying growth of entanglement, including in Floquet systems~\cite{qpsz-pvxm,osipov2026localcorrelationslongrangedualunitary}.
More broadly, similar mechanisms may be explored in frustrated quantum magnets \cite{MiyaharaUeda2003, Balents2010}, higher-spin systems \cite{Affleck1989, Sen2005,PhysRevD.110.094506,PhysRevB.88.245121,PhysRevB.106.174419,Schliemann_2012,pathak2026ergodicitybreakingtransitionqubits}, and coupled spin chains \cite{Schulz1996, RouthGhoshKumar2025}, where approximate local quantum numbers can influence the structure and dynamics of many-body states.

\prlsec{Acknowledgments.---}
A.~P. acknowledges the support from the JST LOTUS Programme (Grant No. Z2025L8290004). A.~K.~S. acknowledges support from JSPS KAKENHI Grant Number JP25K00925.
We thank Tanay Pathak, Sumiran Pujari, and Masaki Tezuka for valuable comments.

\bibliography{refs}

\begin{thebibliography}{68}%
\makeatletter
\providecommand \@ifxundefined [1]{%
 \@ifx{#1\undefined}
}%
\providecommand \@ifnum [1]{%
 \ifnum #1\expandafter \@firstoftwo
 \else \expandafter \@secondoftwo
 \fi
}%
\providecommand \@ifx [1]{%
 \ifx #1\expandafter \@firstoftwo
 \else \expandafter \@secondoftwo
 \fi
}%
\providecommand \natexlab [1]{#1}%
\providecommand \enquote  [1]{``#1''}%
\providecommand \bibnamefont  [1]{#1}%
\providecommand \bibfnamefont [1]{#1}%
\providecommand \citenamefont [1]{#1}%
\providecommand \href@noop [0]{\@secondoftwo}%
\providecommand \href [0]{\begingroup \@sanitize@url \@href}%
\providecommand \@href[1]{\@@startlink{#1}\@@href}%
\providecommand \@@href[1]{\endgroup#1\@@endlink}%
\providecommand \@sanitize@url [0]{\catcode `\\12\catcode `\$12\catcode `\&12\catcode `\#12\catcode `\^12\catcode `\_12\catcode `\%12\relax}%
\providecommand \@@startlink[1]{}%
\providecommand \@@endlink[0]{}%
\providecommand \url  [0]{\begingroup\@sanitize@url \@url }%
\providecommand \@url [1]{\endgroup\@href {#1}{\urlprefix }}%
\providecommand \urlprefix  [0]{URL }%
\providecommand \Eprint [0]{\href }%
\providecommand \doibase [0]{https://doi.org/}%
\providecommand \selectlanguage [0]{\@gobble}%
\providecommand \bibinfo  [0]{\@secondoftwo}%
\providecommand \bibfield  [0]{\@secondoftwo}%
\providecommand \translation [1]{[#1]}%
\providecommand \BibitemOpen [0]{}%
\providecommand \bibitemStop [0]{}%
\providecommand \bibitemNoStop [0]{.\EOS\space}%
\providecommand \EOS [0]{\spacefactor3000\relax}%
\providecommand \BibitemShut  [1]{\csname bibitem#1\endcsname}%
\let\auto@bib@innerbib\@empty
\bibitem [{\citenamefont {Nandkishore}\ and\ \citenamefont {Huse}(2015)}]{NandkishoreHuse2015}%
  \BibitemOpen
  \bibfield  {author} {\bibinfo {author} {\bibfnamefont {R.}~\bibnamefont {Nandkishore}}\ and\ \bibinfo {author} {\bibfnamefont {D.~A.}\ \bibnamefont {Huse}},\ }\bibfield  {title} {\bibinfo {title} {Many-body localization and thermalization in quantum statistical mechanics},\ }\href {https://doi.org/10.1146/annurev-conmatphys-031214-014726} {\bibfield  {journal} {\bibinfo  {journal} {Annual Review of Condensed Matter Physics}\ }\textbf {\bibinfo {volume} {6}},\ \bibinfo {pages} {15} (\bibinfo {year} {2015})}\BibitemShut {NoStop}%
\bibitem [{\citenamefont {Gogolin}\ and\ \citenamefont {Eisert}(2016)}]{GogolinEisert2016}%
  \BibitemOpen
  \bibfield  {author} {\bibinfo {author} {\bibfnamefont {C.}~\bibnamefont {Gogolin}}\ and\ \bibinfo {author} {\bibfnamefont {J.}~\bibnamefont {Eisert}},\ }\bibfield  {title} {\bibinfo {title} {Equilibration, thermalisation, and the emergence of statistical mechanics in closed quantum systems},\ }\href {https://doi.org/10.1088/0034-4885/79/5/056001} {\bibfield  {journal} {\bibinfo  {journal} {Reports on Progress in Physics}\ }\textbf {\bibinfo {volume} {79}},\ \bibinfo {pages} {056001} (\bibinfo {year} {2016})}\BibitemShut {NoStop}%
\bibitem [{\citenamefont {Deutsch}(1991)}]{Deutsch1991}%
  \BibitemOpen
  \bibfield  {author} {\bibinfo {author} {\bibfnamefont {J.~M.}\ \bibnamefont {Deutsch}},\ }\bibfield  {title} {\bibinfo {title} {Quantum statistical mechanics in a closed system},\ }\href {https://doi.org/10.1103/PhysRevA.43.2046} {\bibfield  {journal} {\bibinfo  {journal} {Physical Review A}\ }\textbf {\bibinfo {volume} {43}},\ \bibinfo {pages} {2046} (\bibinfo {year} {1991})}\BibitemShut {NoStop}%
\bibitem [{\citenamefont {Srednicki}(1994)}]{Srednicki}%
  \BibitemOpen
  \bibfield  {author} {\bibinfo {author} {\bibfnamefont {M.}~\bibnamefont {Srednicki}},\ }\bibfield  {title} {\bibinfo {title} {Chaos and quantum thermalization},\ }\href {https://doi.org/10.1103/PhysRevE.50.888} {\bibfield  {journal} {\bibinfo  {journal} {Physical Review E}\ }\textbf {\bibinfo {volume} {50}},\ \bibinfo {pages} {888} (\bibinfo {year} {1994})}\BibitemShut {NoStop}%
\bibitem [{\citenamefont {Rigol}\ \emph {et~al.}(2008)\citenamefont {Rigol}, \citenamefont {Dunjko},\ and\ \citenamefont {Olshanii}}]{RigolNature}%
  \BibitemOpen
  \bibfield  {author} {\bibinfo {author} {\bibfnamefont {M.}~\bibnamefont {Rigol}}, \bibinfo {author} {\bibfnamefont {V.}~\bibnamefont {Dunjko}},\ and\ \bibinfo {author} {\bibfnamefont {M.}~\bibnamefont {Olshanii}},\ }\bibfield  {title} {\bibinfo {title} {Thermalization and its mechanism for generic isolated quantum systems},\ }\href {https://doi.org/10.1038/nature06838} {\bibfield  {journal} {\bibinfo  {journal} {Nature}\ }\textbf {\bibinfo {volume} {452}},\ \bibinfo {pages} {854} (\bibinfo {year} {2008})}\BibitemShut {NoStop}%
\bibitem [{\citenamefont {Deutsch}(2018)}]{Deutsch2018}%
  \BibitemOpen
  \bibfield  {author} {\bibinfo {author} {\bibfnamefont {J.~M.}\ \bibnamefont {Deutsch}},\ }\bibfield  {title} {\bibinfo {title} {Eigenstate thermalization hypothesis},\ }\href {https://doi.org/10.1088/1361-6633/aac9f1} {\bibfield  {journal} {\bibinfo  {journal} {Reports on Progress in Physics}\ }\textbf {\bibinfo {volume} {81}},\ \bibinfo {pages} {082001} (\bibinfo {year} {2018})}\BibitemShut {NoStop}%
\bibitem [{\citenamefont {D'Alessio}\ \emph {et~al.}(2016)\citenamefont {D'Alessio}, \citenamefont {Kafri}, \citenamefont {Polkovnikov},\ and\ \citenamefont {Rigol}}]{DAlessio2016}%
  \BibitemOpen
  \bibfield  {author} {\bibinfo {author} {\bibfnamefont {L.}~\bibnamefont {D'Alessio}}, \bibinfo {author} {\bibfnamefont {Y.}~\bibnamefont {Kafri}}, \bibinfo {author} {\bibfnamefont {A.}~\bibnamefont {Polkovnikov}},\ and\ \bibinfo {author} {\bibfnamefont {M.}~\bibnamefont {Rigol}},\ }\bibfield  {title} {\bibinfo {title} {From quantum chaos and eigenstate thermalization to statistical mechanics and thermodynamics},\ }\href {https://doi.org/10.1080/00018732.2016.1198134} {\bibfield  {journal} {\bibinfo  {journal} {Advances in Physics}\ }\textbf {\bibinfo {volume} {65}},\ \bibinfo {pages} {239} (\bibinfo {year} {2016})}\BibitemShut {NoStop}%
\bibitem [{\citenamefont {Mierzejewski}\ and\ \citenamefont {Vidmar}(2020)}]{MierzejewskiVidmar2020}%
  \BibitemOpen
  \bibfield  {author} {\bibinfo {author} {\bibfnamefont {M.}~\bibnamefont {Mierzejewski}}\ and\ \bibinfo {author} {\bibfnamefont {L.}~\bibnamefont {Vidmar}},\ }\bibfield  {title} {\bibinfo {title} {Quantitative impact of integrals of motion on the eigenstate thermalization hypothesis},\ }\href {https://doi.org/10.1103/PhysRevLett.124.040603} {\bibfield  {journal} {\bibinfo  {journal} {Physical Review Letters}\ }\textbf {\bibinfo {volume} {124}},\ \bibinfo {pages} {040603} (\bibinfo {year} {2020})}\BibitemShut {NoStop}%
\bibitem [{\citenamefont {Oganesyan}\ and\ \citenamefont {Huse}(2007)}]{OganesyanHuse2007}%
  \BibitemOpen
  \bibfield  {author} {\bibinfo {author} {\bibfnamefont {V.}~\bibnamefont {Oganesyan}}\ and\ \bibinfo {author} {\bibfnamefont {D.~A.}\ \bibnamefont {Huse}},\ }\bibfield  {title} {\bibinfo {title} {Localization of interacting fermions at high temperature},\ }\href {https://doi.org/10.1103/PhysRevB.75.155111} {\bibfield  {journal} {\bibinfo  {journal} {Physical Review B}\ }\textbf {\bibinfo {volume} {75}},\ \bibinfo {pages} {155111} (\bibinfo {year} {2007})}\BibitemShut {NoStop}%
\bibitem [{\citenamefont {Atas}\ \emph {et~al.}(2013)\citenamefont {Atas}, \citenamefont {Bogomolny}, \citenamefont {Giraud},\ and\ \citenamefont {Roux}}]{Atas2013}%
  \BibitemOpen
  \bibfield  {author} {\bibinfo {author} {\bibfnamefont {Y.~Y.}\ \bibnamefont {Atas}}, \bibinfo {author} {\bibfnamefont {E.}~\bibnamefont {Bogomolny}}, \bibinfo {author} {\bibfnamefont {O.}~\bibnamefont {Giraud}},\ and\ \bibinfo {author} {\bibfnamefont {G.}~\bibnamefont {Roux}},\ }\bibfield  {title} {\bibinfo {title} {Distribution of the ratio of consecutive level spacings in random matrix ensembles},\ }\href {https://doi.org/10.1103/PhysRevLett.110.084101} {\bibfield  {journal} {\bibinfo  {journal} {Physical Review Letters}\ }\textbf {\bibinfo {volume} {110}},\ \bibinfo {pages} {084101} (\bibinfo {year} {2013})}\BibitemShut {NoStop}%
\bibitem [{\citenamefont {Bohigas}\ \emph {et~al.}(1984)\citenamefont {Bohigas}, \citenamefont {Giannoni},\ and\ \citenamefont {Schmit}}]{BohigasGiannoniSchmit1984}%
  \BibitemOpen
  \bibfield  {author} {\bibinfo {author} {\bibfnamefont {O.}~\bibnamefont {Bohigas}}, \bibinfo {author} {\bibfnamefont {M.~J.}\ \bibnamefont {Giannoni}},\ and\ \bibinfo {author} {\bibfnamefont {C.}~\bibnamefont {Schmit}},\ }\bibfield  {title} {\bibinfo {title} {Characterization of chaotic quantum spectra and universality of level fluctuation laws},\ }\href {https://doi.org/10.1103/PhysRevLett.52.1} {\bibfield  {journal} {\bibinfo  {journal} {Physical Review Letters}\ }\textbf {\bibinfo {volume} {52}},\ \bibinfo {pages} {1} (\bibinfo {year} {1984})}\BibitemShut {NoStop}%
\bibitem [{\citenamefont {Santos}\ and\ \citenamefont {Rigol}(2010{\natexlab{a}})}]{SantosRigol2010}%
  \BibitemOpen
  \bibfield  {author} {\bibinfo {author} {\bibfnamefont {L.~F.}\ \bibnamefont {Santos}}\ and\ \bibinfo {author} {\bibfnamefont {M.}~\bibnamefont {Rigol}},\ }\bibfield  {title} {\bibinfo {title} {Onset of quantum chaos in one-dimensional bosonic and fermionic systems and its relation to thermalization},\ }\href {https://doi.org/10.1103/PhysRevE.81.036206} {\bibfield  {journal} {\bibinfo  {journal} {Physical Review E}\ }\textbf {\bibinfo {volume} {81}},\ \bibinfo {pages} {036206} (\bibinfo {year} {2010}{\natexlab{a}})}\BibitemShut {NoStop}%
\bibitem [{\citenamefont {Pal}\ and\ \citenamefont {Huse}(2010)}]{PalHuse2010}%
  \BibitemOpen
  \bibfield  {author} {\bibinfo {author} {\bibfnamefont {A.}~\bibnamefont {Pal}}\ and\ \bibinfo {author} {\bibfnamefont {D.~A.}\ \bibnamefont {Huse}},\ }\bibfield  {title} {\bibinfo {title} {Many-body localization phase transition},\ }\href {https://doi.org/10.1103/PhysRevB.82.174411} {\bibfield  {journal} {\bibinfo  {journal} {Physical Review B}\ }\textbf {\bibinfo {volume} {82}},\ \bibinfo {pages} {174411} (\bibinfo {year} {2010})}\BibitemShut {NoStop}%
\bibitem [{\citenamefont {LeBlond}\ \emph {et~al.}(2021)\citenamefont {LeBlond}, \citenamefont {Sels}, \citenamefont {Polkovnikov},\ and\ \citenamefont {Rigol}}]{LeBlondSelsPolkovnikovRigol2021}%
  \BibitemOpen
  \bibfield  {author} {\bibinfo {author} {\bibfnamefont {T.}~\bibnamefont {LeBlond}}, \bibinfo {author} {\bibfnamefont {D.}~\bibnamefont {Sels}}, \bibinfo {author} {\bibfnamefont {A.}~\bibnamefont {Polkovnikov}},\ and\ \bibinfo {author} {\bibfnamefont {M.}~\bibnamefont {Rigol}},\ }\bibfield  {title} {\bibinfo {title} {Universality in the onset of quantum chaos in many-body systems},\ }\href {https://doi.org/10.1103/PhysRevB.104.L201117} {\bibfield  {journal} {\bibinfo  {journal} {Physical Review B}\ }\textbf {\bibinfo {volume} {104}},\ \bibinfo {pages} {L201117} (\bibinfo {year} {2021})}\BibitemShut {NoStop}%
\bibitem [{\citenamefont {Richter}\ \emph {et~al.}(2020)\citenamefont {Richter}, \citenamefont {Dymarsky}, \citenamefont {Steinigeweg},\ and\ \citenamefont {Gemmer}}]{RichterDymarskySteinigewegGemmer2020}%
  \BibitemOpen
  \bibfield  {author} {\bibinfo {author} {\bibfnamefont {J.}~\bibnamefont {Richter}}, \bibinfo {author} {\bibfnamefont {A.}~\bibnamefont {Dymarsky}}, \bibinfo {author} {\bibfnamefont {R.}~\bibnamefont {Steinigeweg}},\ and\ \bibinfo {author} {\bibfnamefont {J.}~\bibnamefont {Gemmer}},\ }\bibfield  {title} {\bibinfo {title} {Eigenstate thermalization hypothesis beyond standard indicators: Emergence of random-matrix behavior at small frequencies},\ }\href {https://doi.org/10.1103/PhysRevE.102.042127} {\bibfield  {journal} {\bibinfo  {journal} {Physical Review E}\ }\textbf {\bibinfo {volume} {102}},\ \bibinfo {pages} {042127} (\bibinfo {year} {2020})}\BibitemShut {NoStop}%
\bibitem [{\citenamefont {Knetter}\ \emph {et~al.}(2003)\citenamefont {Knetter}, \citenamefont {Schmidt},\ and\ \citenamefont {Uhrig}}]{KnetterSchmidtUhrig2003}%
  \BibitemOpen
  \bibfield  {author} {\bibinfo {author} {\bibfnamefont {C.}~\bibnamefont {Knetter}}, \bibinfo {author} {\bibfnamefont {K.~P.}\ \bibnamefont {Schmidt}},\ and\ \bibinfo {author} {\bibfnamefont {G.~S.}\ \bibnamefont {Uhrig}},\ }\bibfield  {title} {\bibinfo {title} {High order perturbation theory for spectral densities of multi-particle excitations: {$S=1/2$} two-leg {Heisenberg} ladder},\ }\href {https://doi.org/10.1140/epjb/e2004-00008-2} {\bibfield  {journal} {\bibinfo  {journal} {The European Physical Journal B}\ }\textbf {\bibinfo {volume} {36}},\ \bibinfo {pages} {525} (\bibinfo {year} {2003})}\BibitemShut {NoStop}%
\bibitem [{\citenamefont {Schmiedinghoff}\ \emph {et~al.}(2022)\citenamefont {Schmiedinghoff}, \citenamefont {M{\"u}ller}, \citenamefont {Kumar}, \citenamefont {Uhrig},\ and\ \citenamefont {Fauseweh}}]{SchmiedinghoffEtAl2022}%
  \BibitemOpen
  \bibfield  {author} {\bibinfo {author} {\bibfnamefont {G.}~\bibnamefont {Schmiedinghoff}}, \bibinfo {author} {\bibfnamefont {L.}~\bibnamefont {M{\"u}ller}}, \bibinfo {author} {\bibfnamefont {U.}~\bibnamefont {Kumar}}, \bibinfo {author} {\bibfnamefont {G.~S.}\ \bibnamefont {Uhrig}},\ and\ \bibinfo {author} {\bibfnamefont {B.}~\bibnamefont {Fauseweh}},\ }\bibfield  {title} {\bibinfo {title} {Three-body bound states in antiferromagnetic spin ladders},\ }\href {https://doi.org/10.1038/s42005-022-00986-0} {\bibfield  {journal} {\bibinfo  {journal} {Communications Physics}\ }\textbf {\bibinfo {volume} {5}},\ \bibinfo {pages} {218} (\bibinfo {year} {2022})}\BibitemShut {NoStop}%
\bibitem [{\citenamefont {Wildeboer}\ \emph {et~al.}(2026)\citenamefont {Wildeboer}, \citenamefont {Lajer},\ and\ \citenamefont {Konik}}]{WildeboerLajerKonik2026}%
  \BibitemOpen
  \bibfield  {author} {\bibinfo {author} {\bibfnamefont {J.}~\bibnamefont {Wildeboer}}, \bibinfo {author} {\bibfnamefont {M.}~\bibnamefont {Lajer}},\ and\ \bibinfo {author} {\bibfnamefont {R.~M.}\ \bibnamefont {Konik}},\ }\bibfield  {title} {\bibinfo {title} {Breakdown of quantum chaos in the staggered-field {XXZ} chain: Confinement and meson formation},\ }\href {https://doi.org/10.1103/t26r-gz35} {\bibfield  {journal} {\bibinfo  {journal} {Physical Review B}\ }\textbf {\bibinfo {volume} {114}},\ \bibinfo {pages} {214428} (\bibinfo {year} {2026})}\BibitemShut {NoStop}%
\bibitem [{\citenamefont {Bulchandani}\ \emph {et~al.}(2022)\citenamefont {Bulchandani}, \citenamefont {Huse},\ and\ \citenamefont {Gopalakrishnan}}]{BulchandaniHuseGopalakrishnan2022}%
  \BibitemOpen
  \bibfield  {author} {\bibinfo {author} {\bibfnamefont {V.~B.}\ \bibnamefont {Bulchandani}}, \bibinfo {author} {\bibfnamefont {D.~A.}\ \bibnamefont {Huse}},\ and\ \bibinfo {author} {\bibfnamefont {S.}~\bibnamefont {Gopalakrishnan}},\ }\bibfield  {title} {\bibinfo {title} {Onset of many-body quantum chaos due to breaking integrability},\ }\href {https://doi.org/10.1103/PhysRevB.105.214308} {\bibfield  {journal} {\bibinfo  {journal} {Physical Review B}\ }\textbf {\bibinfo {volume} {105}},\ \bibinfo {pages} {214308} (\bibinfo {year} {2022})}\BibitemShut {NoStop}%
\bibitem [{\citenamefont {Birnkammer}\ \emph {et~al.}(2022)\citenamefont {Birnkammer}, \citenamefont {Bastianello},\ and\ \citenamefont {Knap}}]{BirnkammerBastianelloKnap2022}%
  \BibitemOpen
  \bibfield  {author} {\bibinfo {author} {\bibfnamefont {S.}~\bibnamefont {Birnkammer}}, \bibinfo {author} {\bibfnamefont {A.}~\bibnamefont {Bastianello}},\ and\ \bibinfo {author} {\bibfnamefont {M.}~\bibnamefont {Knap}},\ }\bibfield  {title} {\bibinfo {title} {Prethermalization in one-dimensional quantum many-body systems with confinement},\ }\href {https://doi.org/10.1038/s41467-022-35301-6} {\bibfield  {journal} {\bibinfo  {journal} {Nature Communications}\ }\textbf {\bibinfo {volume} {13}},\ \bibinfo {pages} {7663} (\bibinfo {year} {2022})}\BibitemShut {NoStop}%
\bibitem [{\citenamefont {Paw{\l}owski}\ \emph {et~al.}(2024)\citenamefont {Paw{\l}owski}, \citenamefont {Panfil}, \citenamefont {Herbrych},\ and\ \citenamefont {Mierzejewski}}]{PawlowskiPanfilHerbrychMierzejewski2024}%
  \BibitemOpen
  \bibfield  {author} {\bibinfo {author} {\bibfnamefont {J.}~\bibnamefont {Paw{\l}owski}}, \bibinfo {author} {\bibfnamefont {M.}~\bibnamefont {Panfil}}, \bibinfo {author} {\bibfnamefont {J.}~\bibnamefont {Herbrych}},\ and\ \bibinfo {author} {\bibfnamefont {M.}~\bibnamefont {Mierzejewski}},\ }\bibfield  {title} {\bibinfo {title} {Long-living prethermalization in nearly integrable spin ladders},\ }\href {https://doi.org/10.1103/PhysRevB.109.L161109} {\bibfield  {journal} {\bibinfo  {journal} {Physical Review B}\ }\textbf {\bibinfo {volume} {109}},\ \bibinfo {pages} {L161109} (\bibinfo {year} {2024})}\BibitemShut {NoStop}%
\bibitem [{\citenamefont {Mallayya}\ \emph {et~al.}(2019)\citenamefont {Mallayya}, \citenamefont {Rigol},\ and\ \citenamefont {Roeck}}]{MallayyaRigolDeRoeck2019}%
  \BibitemOpen
  \bibfield  {author} {\bibinfo {author} {\bibfnamefont {K.}~\bibnamefont {Mallayya}}, \bibinfo {author} {\bibfnamefont {M.}~\bibnamefont {Rigol}},\ and\ \bibinfo {author} {\bibfnamefont {W.~D.}\ \bibnamefont {Roeck}},\ }\bibfield  {title} {\bibinfo {title} {Prethermalization and thermalization in isolated quantum systems},\ }\href {https://doi.org/10.1103/PhysRevX.9.021027} {\bibfield  {journal} {\bibinfo  {journal} {Physical Review X}\ }\textbf {\bibinfo {volume} {9}},\ \bibinfo {pages} {021027} (\bibinfo {year} {2019})}\BibitemShut {NoStop}%
\bibitem [{\citenamefont {Barnes}\ \emph {et~al.}(1993)\citenamefont {Barnes}, \citenamefont {Dagotto}, \citenamefont {Riera},\ and\ \citenamefont {Swanson}}]{BarnesEtAl1993}%
  \BibitemOpen
  \bibfield  {author} {\bibinfo {author} {\bibfnamefont {T.}~\bibnamefont {Barnes}}, \bibinfo {author} {\bibfnamefont {E.}~\bibnamefont {Dagotto}}, \bibinfo {author} {\bibfnamefont {J.}~\bibnamefont {Riera}},\ and\ \bibinfo {author} {\bibfnamefont {E.~S.}\ \bibnamefont {Swanson}},\ }\bibfield  {title} {\bibinfo {title} {Excitation spectrum of {Heisenberg} spin ladders},\ }\href {https://doi.org/10.1103/PhysRevB.47.3196} {\bibfield  {journal} {\bibinfo  {journal} {Physical Review B}\ }\textbf {\bibinfo {volume} {47}},\ \bibinfo {pages} {3196} (\bibinfo {year} {1993})}\BibitemShut {NoStop}%
\bibitem [{\citenamefont {Dagotto}\ and\ \citenamefont {Rice}(1996)}]{DagottoRice1996}%
  \BibitemOpen
  \bibfield  {author} {\bibinfo {author} {\bibfnamefont {E.}~\bibnamefont {Dagotto}}\ and\ \bibinfo {author} {\bibfnamefont {T.~M.}\ \bibnamefont {Rice}},\ }\bibfield  {title} {\bibinfo {title} {Surprises on the way from one- to two-dimensional quantum magnets: The ladder materials},\ }\href {https://doi.org/10.1126/science.271.5249.618} {\bibfield  {journal} {\bibinfo  {journal} {Science}\ }\textbf {\bibinfo {volume} {271}},\ \bibinfo {pages} {618} (\bibinfo {year} {1996})}\BibitemShut {NoStop}%
\bibitem [{\citenamefont {Shelton}\ \emph {et~al.}(1996)\citenamefont {Shelton}, \citenamefont {Nersesyan},\ and\ \citenamefont {Tsvelik}}]{PhysRevB.53.8521}%
  \BibitemOpen
  \bibfield  {author} {\bibinfo {author} {\bibfnamefont {D.~G.}\ \bibnamefont {Shelton}}, \bibinfo {author} {\bibfnamefont {A.~A.}\ \bibnamefont {Nersesyan}},\ and\ \bibinfo {author} {\bibfnamefont {A.~M.}\ \bibnamefont {Tsvelik}},\ }\bibfield  {title} {\bibinfo {title} {Antiferromagnetic spin ladders: Crossover between spin {S}=1/2 and {S}=1 chains},\ }\href {https://doi.org/10.1103/PhysRevB.53.8521} {\bibfield  {journal} {\bibinfo  {journal} {Physical Review B}\ }\textbf {\bibinfo {volume} {53}},\ \bibinfo {pages} {8521} (\bibinfo {year} {1996})}\BibitemShut {NoStop}%
\bibitem [{\citenamefont {Khodja}\ \emph {et~al.}(2016)\citenamefont {Khodja}, \citenamefont {Schmidtke},\ and\ \citenamefont {Gemmer}}]{PhysRevE.93.042101}%
  \BibitemOpen
  \bibfield  {author} {\bibinfo {author} {\bibfnamefont {A.}~\bibnamefont {Khodja}}, \bibinfo {author} {\bibfnamefont {D.}~\bibnamefont {Schmidtke}},\ and\ \bibinfo {author} {\bibfnamefont {J.}~\bibnamefont {Gemmer}},\ }\bibfield  {title} {\bibinfo {title} {Initial-state-independent equilibration at the breakdown of the eigenstate thermalization hypothesis},\ }\href {https://doi.org/10.1103/PhysRevE.93.042101} {\bibfield  {journal} {\bibinfo  {journal} {Phys. Rev. E}\ }\textbf {\bibinfo {volume} {93}},\ \bibinfo {pages} {042101} (\bibinfo {year} {2016})}\BibitemShut {NoStop}%
\bibitem [{\citenamefont {{\ifmmode \mbox{\c{C}}\else \c{C}\fi{}}even}\ \emph {et~al.}(2026)\citenamefont {{\ifmmode \mbox{\c{C}}\else \c{C}\fi{}}even}, \citenamefont {Peinemann},\ and\ \citenamefont {Heidrich-Meisner}}]{85kv-972r}%
  \BibitemOpen
  \bibfield  {author} {\bibinfo {author} {\bibfnamefont {K.}~\bibnamefont {{\ifmmode \mbox{\c{C}}\else \c{C}\fi{}}even}}, \bibinfo {author} {\bibfnamefont {L.}~\bibnamefont {Peinemann}},\ and\ \bibinfo {author} {\bibfnamefont {F.}~\bibnamefont {Heidrich-Meisner}},\ }\bibfield  {title} {\bibinfo {title} {Hierarchy of timescales in a disordered spin-$\frac{1}{2}$ {XX} ladder},\ }\href {https://doi.org/10.1103/85kv-972r} {\bibfield  {journal} {\bibinfo  {journal} {Phys. Rev. B}\ }\textbf {\bibinfo {volume} {113}},\ \bibinfo {pages} {045126} (\bibinfo {year} {2026})}\BibitemShut {NoStop}%
\bibitem [{\citenamefont {McClarty}\ \emph {et~al.}(2020)\citenamefont {McClarty}, \citenamefont {Haque}, \citenamefont {Sen},\ and\ \citenamefont {Richter}}]{McClarty2020}%
  \BibitemOpen
  \bibfield  {author} {\bibinfo {author} {\bibfnamefont {P.~A.}\ \bibnamefont {McClarty}}, \bibinfo {author} {\bibfnamefont {M.}~\bibnamefont {Haque}}, \bibinfo {author} {\bibfnamefont {A.}~\bibnamefont {Sen}},\ and\ \bibinfo {author} {\bibfnamefont {J.}~\bibnamefont {Richter}},\ }\bibfield  {title} {\bibinfo {title} {Disorder-free localization and many-body quantum scars from magnetic frustration},\ }\href {https://doi.org/10.1103/PhysRevB.102.224303} {\bibfield  {journal} {\bibinfo  {journal} {Physical Review B}\ }\textbf {\bibinfo {volume} {102}},\ \bibinfo {pages} {224303} (\bibinfo {year} {2020})}\BibitemShut {NoStop}%
\bibitem [{\citenamefont {Iadecola}\ and\ \citenamefont {{\v Z}nidari{\v c}}(2019)}]{IadecolaZnidaric2019}%
  \BibitemOpen
  \bibfield  {author} {\bibinfo {author} {\bibfnamefont {T.}~\bibnamefont {Iadecola}}\ and\ \bibinfo {author} {\bibfnamefont {M.}~\bibnamefont {{\v Z}nidari{\v c}}},\ }\bibfield  {title} {\bibinfo {title} {Exact localized and ballistic eigenstates in disordered chaotic spin ladders and the {Fermi-Hubbard} model},\ }\href {https://doi.org/10.1103/PhysRevLett.123.036403} {\bibfield  {journal} {\bibinfo  {journal} {Physical Review Letters}\ }\textbf {\bibinfo {volume} {123}},\ \bibinfo {pages} {036403} (\bibinfo {year} {2019})}\BibitemShut {NoStop}%
\bibitem [{\citenamefont {Najafzadeh}\ and\ \citenamefont {Langari}(2026)}]{NajafzadehLangari2026}%
  \BibitemOpen
  \bibfield  {author} {\bibinfo {author} {\bibfnamefont {H.}~\bibnamefont {Najafzadeh}}\ and\ \bibinfo {author} {\bibfnamefont {A.}~\bibnamefont {Langari}},\ }\bibfield  {title} {\bibinfo {title} {Crossover from quantum chaos to a reversed quantum disentangled liquid in a disorder-free spin ladder},\ }\href {https://doi.org/10.1103/12pf-d8p8} {\bibfield  {journal} {\bibinfo  {journal} {Physical Review B}\ }\textbf {\bibinfo {volume} {113}},\ \bibinfo {pages} {144308} (\bibinfo {year} {2026})}\BibitemShut {NoStop}%
\bibitem [{\citenamefont {{\ifmmode \check{Z}\else \v{Z}\fi{}}nidari{\ifmmode \check{c}\else \v{c}\fi{}}}(2013{\natexlab{a}})}]{PhysRevLett.110.070602}%
  \BibitemOpen
  \bibfield  {author} {\bibinfo {author} {\bibfnamefont {M.}~\bibnamefont {{\ifmmode \check{Z}\else \v{Z}\fi{}}nidari{\ifmmode \check{c}\else \v{c}\fi{}}}},\ }\bibfield  {title} {\bibinfo {title} {Coexistence of diffusive and ballistic transport in a simple spin ladder},\ }\href {https://doi.org/10.1103/PhysRevLett.110.070602} {\bibfield  {journal} {\bibinfo  {journal} {Phys. Rev. Lett.}\ }\textbf {\bibinfo {volume} {110}},\ \bibinfo {pages} {070602} (\bibinfo {year} {2013}{\natexlab{a}})}\BibitemShut {NoStop}%
\bibitem [{\citenamefont {{\ifmmode \check{Z}\else \v{Z}\fi{}}nidari{\ifmmode \check{c}\else \v{c}\fi{}}}(2013{\natexlab{b}})}]{PhysRevB.88.205135}%
  \BibitemOpen
  \bibfield  {author} {\bibinfo {author} {\bibfnamefont {M.}~\bibnamefont {{\ifmmode \check{Z}\else \v{Z}\fi{}}nidari{\ifmmode \check{c}\else \v{c}\fi{}}}},\ }\bibfield  {title} {\bibinfo {title} {Magnetization transport in spin ladders and next-nearest-neighbor chains},\ }\href {https://doi.org/10.1103/PhysRevB.88.205135} {\bibfield  {journal} {\bibinfo  {journal} {Phys. Rev. B}\ }\textbf {\bibinfo {volume} {88}},\ \bibinfo {pages} {205135} (\bibinfo {year} {2013}{\natexlab{b}})}\BibitemShut {NoStop}%
\bibitem [{\citenamefont {Sachdev}\ and\ \citenamefont {Bhatt}(1990)}]{SachdevBhatt1990}%
  \BibitemOpen
  \bibfield  {author} {\bibinfo {author} {\bibfnamefont {S.}~\bibnamefont {Sachdev}}\ and\ \bibinfo {author} {\bibfnamefont {R.~N.}\ \bibnamefont {Bhatt}},\ }\bibfield  {title} {\bibinfo {title} {Bond-operator representation of quantum spins: Mean-field theory of frustrated quantum {Heisenberg} antiferromagnets},\ }\href {https://doi.org/10.1103/PhysRevB.41.9323} {\bibfield  {journal} {\bibinfo  {journal} {Physical Review B}\ }\textbf {\bibinfo {volume} {41}},\ \bibinfo {pages} {9323} (\bibinfo {year} {1990})}\BibitemShut {NoStop}%
\bibitem [{\citenamefont {Gopalan}\ \emph {et~al.}(1994)\citenamefont {Gopalan}, \citenamefont {Rice},\ and\ \citenamefont {Sigrist}}]{GopalanRiceSigrist1994}%
  \BibitemOpen
  \bibfield  {author} {\bibinfo {author} {\bibfnamefont {S.}~\bibnamefont {Gopalan}}, \bibinfo {author} {\bibfnamefont {T.~M.}\ \bibnamefont {Rice}},\ and\ \bibinfo {author} {\bibfnamefont {M.}~\bibnamefont {Sigrist}},\ }\bibfield  {title} {\bibinfo {title} {Spin ladders with spin gaps: A description of a class of cuprates},\ }\href {https://doi.org/10.1103/PhysRevB.49.8901} {\bibfield  {journal} {\bibinfo  {journal} {Physical Review B}\ }\textbf {\bibinfo {volume} {49}},\ \bibinfo {pages} {8901} (\bibinfo {year} {1994})}\BibitemShut {NoStop}%
\bibitem [{\citenamefont {Knetter}\ \emph {et~al.}(2001)\citenamefont {Knetter}, \citenamefont {Schmidt}, \citenamefont {Gr{\"u}ninger},\ and\ \citenamefont {Uhrig}}]{KnetterSchmidtGruningerUhrig2001}%
  \BibitemOpen
  \bibfield  {author} {\bibinfo {author} {\bibfnamefont {C.}~\bibnamefont {Knetter}}, \bibinfo {author} {\bibfnamefont {K.~P.}\ \bibnamefont {Schmidt}}, \bibinfo {author} {\bibfnamefont {M.}~\bibnamefont {Gr{\"u}ninger}},\ and\ \bibinfo {author} {\bibfnamefont {G.~S.}\ \bibnamefont {Uhrig}},\ }\bibfield  {title} {\bibinfo {title} {Fractional and integer excitations in quantum antiferromagnetic spin-1/2 ladders},\ }\href {https://doi.org/10.1103/PhysRevLett.87.167204} {\bibfield  {journal} {\bibinfo  {journal} {Physical Review Letters}\ }\textbf {\bibinfo {volume} {87}},\ \bibinfo {pages} {167204} (\bibinfo {year} {2001})}\BibitemShut {NoStop}%
\bibitem [{\citenamefont {Baygan}\ \emph {et~al.}(2015)\citenamefont {Baygan}, \citenamefont {Lim},\ and\ \citenamefont {Sheng}}]{BayganLimSheng2015}%
  \BibitemOpen
  \bibfield  {author} {\bibinfo {author} {\bibfnamefont {E.}~\bibnamefont {Baygan}}, \bibinfo {author} {\bibfnamefont {S.~P.}\ \bibnamefont {Lim}},\ and\ \bibinfo {author} {\bibfnamefont {D.~N.}\ \bibnamefont {Sheng}},\ }\bibfield  {title} {\bibinfo {title} {Many-body localization and mobility edge in a disordered spin-1/2 {Heisenberg} ladder},\ }\href {https://doi.org/10.1103/PhysRevB.92.195153} {\bibfield  {journal} {\bibinfo  {journal} {Physical Review B}\ }\textbf {\bibinfo {volume} {92}},\ \bibinfo {pages} {195153} (\bibinfo {year} {2015})}\BibitemShut {NoStop}%
\bibitem [{\citenamefont {Hu}\ \emph {et~al.}(2026)\citenamefont {Hu}, \citenamefont {Hong}, \citenamefont {Guo},\ and\ \citenamefont {Ren}}]{HuHongGuoRen2026}%
  \BibitemOpen
  \bibfield  {author} {\bibinfo {author} {\bibfnamefont {T.}~\bibnamefont {Hu}}, \bibinfo {author} {\bibfnamefont {J.}~\bibnamefont {Hong}}, \bibinfo {author} {\bibfnamefont {D.}~\bibnamefont {Guo}},\ and\ \bibinfo {author} {\bibfnamefont {H.}~\bibnamefont {Ren}},\ }\bibfield  {title} {\bibinfo {title} {Many-body localization properties of fully frustrated {Heisenberg} spin-1/2 ladder model with next-nearest-neighbor interaction},\ }\href {https://doi.org/10.1016/j.physa.2026.131303} {\bibfield  {journal} {\bibinfo  {journal} {Physica A: Statistical Mechanics and its Applications}\ }\textbf {\bibinfo {volume} {685}},\ \bibinfo {pages} {131303} (\bibinfo {year} {2026})}\BibitemShut {NoStop}%
\bibitem [{\citenamefont {Richter}\ and\ \citenamefont {Pal}(2022{\natexlab{a}})}]{RichterPal2022}%
  \BibitemOpen
  \bibfield  {author} {\bibinfo {author} {\bibfnamefont {J.}~\bibnamefont {Richter}}\ and\ \bibinfo {author} {\bibfnamefont {A.}~\bibnamefont {Pal}},\ }\bibfield  {title} {\bibinfo {title} {Many-body localization and delocalization dynamics in the thermodynamic limit},\ }\href {https://doi.org/10.1103/PhysRevB.105.L220405} {\bibfield  {journal} {\bibinfo  {journal} {Physical Review B}\ }\textbf {\bibinfo {volume} {105}},\ \bibinfo {pages} {L220405} (\bibinfo {year} {2022}{\natexlab{a}})}\BibitemShut {NoStop}%
\bibitem [{\citenamefont {Pal}\ \emph {et~al.}(2025)\citenamefont {Pal}, \citenamefont {Sarkar}, \citenamefont {Sengupta},\ and\ \citenamefont {Sen}}]{PalEtAl2025}%
  \BibitemOpen
  \bibfield  {author} {\bibinfo {author} {\bibfnamefont {M.}~\bibnamefont {Pal}}, \bibinfo {author} {\bibfnamefont {M.}~\bibnamefont {Sarkar}}, \bibinfo {author} {\bibfnamefont {K.}~\bibnamefont {Sengupta}},\ and\ \bibinfo {author} {\bibfnamefont {A.}~\bibnamefont {Sen}},\ }\bibfield  {title} {\bibinfo {title} {Scar-induced imbalance in staggered {Rydberg} ladders},\ }\href {https://doi.org/10.1103/PhysRevB.111.L161101} {\bibfield  {journal} {\bibinfo  {journal} {Physical Review B}\ }\textbf {\bibinfo {volume} {111}},\ \bibinfo {pages} {L161101} (\bibinfo {year} {2025})}\BibitemShut {NoStop}%
\bibitem [{\citenamefont {Chandran}\ \emph {et~al.}(2023)\citenamefont {Chandran}, \citenamefont {Iadecola}, \citenamefont {Khemani},\ and\ \citenamefont {Moessner}}]{scars_review}%
  \BibitemOpen
  \bibfield  {author} {\bibinfo {author} {\bibfnamefont {A.}~\bibnamefont {Chandran}}, \bibinfo {author} {\bibfnamefont {T.}~\bibnamefont {Iadecola}}, \bibinfo {author} {\bibfnamefont {V.}~\bibnamefont {Khemani}},\ and\ \bibinfo {author} {\bibfnamefont {R.}~\bibnamefont {Moessner}},\ }\bibfield  {title} {\bibinfo {title} {Quantum many-body scars: A quasiparticle perspective},\ }\href {https://doi.org/10.1146/annurev-conmatphys-031620-101617} {\bibfield  {journal} {\bibinfo  {journal} {Annual Review of Condensed Matter Physics}\ }\textbf {\bibinfo {volume} {14}},\ \bibinfo {pages} {443} (\bibinfo {year} {2023})}\BibitemShut {NoStop}%
\bibitem [{\citenamefont {Santos}\ and\ \citenamefont {Rigol}(2010{\natexlab{b}})}]{SantosRigolSymmetries2010}%
  \BibitemOpen
  \bibfield  {author} {\bibinfo {author} {\bibfnamefont {L.~F.}\ \bibnamefont {Santos}}\ and\ \bibinfo {author} {\bibfnamefont {M.}~\bibnamefont {Rigol}},\ }\bibfield  {title} {\bibinfo {title} {Localization and the effects of symmetries in the thermalization properties of one-dimensional quantum systems},\ }\href {https://doi.org/10.1103/PhysRevE.82.031130} {\bibfield  {journal} {\bibinfo  {journal} {Physical Review E}\ }\textbf {\bibinfo {volume} {82}},\ \bibinfo {pages} {031130} (\bibinfo {year} {2010}{\natexlab{b}})}\BibitemShut {NoStop}%
\bibitem [{sup()}]{supp_info}%
  \BibitemOpen
  \href@noop {} {}\bibinfo {note} {See Supplemental Material at URL-will-be-inserted-by-publisher, for details on numerics and additional supporting results.}\BibitemShut {Stop}%
\bibitem [{run()}]{rungnote}%
  \BibitemOpen
  \href@noop {} {}\bibinfo {note} {The local rung states are: $|s\rangle=\frac{1}{\sqrt{2}}(|\uparrow\downarrow\rangle-|\downarrow\uparrow\rangle)$ as the rung singlet and the three triplet states, $|t_+\rangle=|\uparrow\uparrow\rangle$, $|t_0\rangle=\frac{1}{\sqrt{2}}(|\uparrow\downarrow\rangle+|\downarrow\uparrow\rangle)$, and $|t_-\rangle=|\downarrow\downarrow\rangle$, such that $P_i^t|s\rangle=0$ and $P_i^t|t_m\rangle_i=|t_m\rangle_i$ for $m=\{+,0,-\}$}\BibitemShut {NoStop}%
\bibitem [{\citenamefont {Rabson}\ \emph {et~al.}(2004)\citenamefont {Rabson}, \citenamefont {Narozhny},\ and\ \citenamefont {Millis}}]{RabsonNarozhnyMillis2004}%
  \BibitemOpen
  \bibfield  {author} {\bibinfo {author} {\bibfnamefont {D.~A.}\ \bibnamefont {Rabson}}, \bibinfo {author} {\bibfnamefont {B.~N.}\ \bibnamefont {Narozhny}},\ and\ \bibinfo {author} {\bibfnamefont {A.~J.}\ \bibnamefont {Millis}},\ }\bibfield  {title} {\bibinfo {title} {Crossover from {Poisson} to {Wigner-Dyson} level statistics in spin chains with integrability breaking},\ }\href {https://doi.org/10.1103/PhysRevB.69.054403} {\bibfield  {journal} {\bibinfo  {journal} {Physical Review B}\ }\textbf {\bibinfo {volume} {69}},\ \bibinfo {pages} {054403} (\bibinfo {year} {2004})}\BibitemShut {NoStop}%
\bibitem [{\citenamefont {Sz{\'a}sz-Schagrin}\ \emph {et~al.}(2021)\citenamefont {Sz{\'a}sz-Schagrin}, \citenamefont {Pozsgay},\ and\ \citenamefont {Tak{\'a}cs}}]{SzaszSchagrinPozsgayTakacs2021}%
  \BibitemOpen
  \bibfield  {author} {\bibinfo {author} {\bibfnamefont {D.}~\bibnamefont {Sz{\'a}sz-Schagrin}}, \bibinfo {author} {\bibfnamefont {B.}~\bibnamefont {Pozsgay}},\ and\ \bibinfo {author} {\bibfnamefont {G.}~\bibnamefont {Tak{\'a}cs}},\ }\bibfield  {title} {\bibinfo {title} {Weak integrability breaking and level spacing distribution},\ }\href {https://doi.org/10.21468/SciPostPhys.11.2.037} {\bibfield  {journal} {\bibinfo  {journal} {SciPost Physics}\ }\textbf {\bibinfo {volume} {11}},\ \bibinfo {pages} {037} (\bibinfo {year} {2021})}\BibitemShut {NoStop}%
\bibitem [{not()}]{note_triplet_occupation}%
  \BibitemOpen
  \href@noop {} {}\bibinfo {note} {Note that using $(P_i^t)^2=P_i^t$, the exact relation is $\Delta N_{t,\alpha}^2/L=(1/L)\sum_i[\langle P_i^t\rangle_\alpha-\langle P_i^t\rangle_\alpha^2]+\sum_{r=1}^{L-1}C_{tt}^{(r),\rm conn}$, where $C_{tt}^{(r),\rm conn}$ generalizes Eq.~\eqref{eq:Cconn} to rung separation $r$.}\BibitemShut {Stop}%
\bibitem [{\citenamefont {Beugeling}\ \emph {et~al.}(2014)\citenamefont {Beugeling}, \citenamefont {Moessner},\ and\ \citenamefont {Haque}}]{Beugeling_etal_pre_2014}%
  \BibitemOpen
  \bibfield  {author} {\bibinfo {author} {\bibfnamefont {W.}~\bibnamefont {Beugeling}}, \bibinfo {author} {\bibfnamefont {R.}~\bibnamefont {Moessner}},\ and\ \bibinfo {author} {\bibfnamefont {M.}~\bibnamefont {Haque}},\ }\bibfield  {title} {\bibinfo {title} {Finite-size scaling of eigenstate thermalization},\ }\href {https://doi.org/10.1103/PhysRevE.89.042112} {\bibfield  {journal} {\bibinfo  {journal} {Physical Review E}\ }\textbf {\bibinfo {volume} {89}},\ \bibinfo {pages} {042112} (\bibinfo {year} {2014})}\BibitemShut {NoStop}%
\bibitem [{\citenamefont {Srednicki}(1999)}]{Srednicki1999}%
  \BibitemOpen
  \bibfield  {author} {\bibinfo {author} {\bibfnamefont {M.}~\bibnamefont {Srednicki}},\ }\bibfield  {title} {\bibinfo {title} {The approach to thermal equilibrium in quantized chaotic systems},\ }\href {https://doi.org/10.1088/0305-4470/32/7/007} {\bibfield  {journal} {\bibinfo  {journal} {Journal of Physics A}\ }\textbf {\bibinfo {volume} {32}},\ \bibinfo {pages} {1163} (\bibinfo {year} {1999})}\BibitemShut {NoStop}%
\bibitem [{\citenamefont {Page}(1993)}]{Page1993}%
  \BibitemOpen
  \bibfield  {author} {\bibinfo {author} {\bibfnamefont {D.~N.}\ \bibnamefont {Page}},\ }\bibfield  {title} {\bibinfo {title} {Average entropy of a subsystem},\ }\href {https://doi.org/10.1103/PhysRevLett.71.1291} {\bibfield  {journal} {\bibinfo  {journal} {Physical Review Letters}\ }\textbf {\bibinfo {volume} {71}},\ \bibinfo {pages} {1291} (\bibinfo {year} {1993})}\BibitemShut {NoStop}%
\bibitem [{\citenamefont {Vidmar}\ and\ \citenamefont {Rigol}(2017)}]{VidmarRigol2017}%
  \BibitemOpen
  \bibfield  {author} {\bibinfo {author} {\bibfnamefont {L.}~\bibnamefont {Vidmar}}\ and\ \bibinfo {author} {\bibfnamefont {M.}~\bibnamefont {Rigol}},\ }\bibfield  {title} {\bibinfo {title} {Entanglement entropy of eigenstates of quantum chaotic hamiltonians},\ }\href {https://doi.org/10.1103/PhysRevLett.119.220603} {\bibfield  {journal} {\bibinfo  {journal} {Physical Review Letters}\ }\textbf {\bibinfo {volume} {119}},\ \bibinfo {pages} {220603} (\bibinfo {year} {2017})}\BibitemShut {NoStop}%
\bibitem [{\citenamefont {Haque}\ \emph {et~al.}(2022)\citenamefont {Haque}, \citenamefont {McClarty},\ and\ \citenamefont {Khaymovich}}]{HaqueMcClartyKhaymovich2022}%
  \BibitemOpen
  \bibfield  {author} {\bibinfo {author} {\bibfnamefont {M.}~\bibnamefont {Haque}}, \bibinfo {author} {\bibfnamefont {P.~A.}\ \bibnamefont {McClarty}},\ and\ \bibinfo {author} {\bibfnamefont {I.~M.}\ \bibnamefont {Khaymovich}},\ }\bibfield  {title} {\bibinfo {title} {Entanglement of midspectrum eigenstates of chaotic many-body systems: Reasons for deviation from random ensembles},\ }\href {https://doi.org/10.1103/PhysRevE.105.014109} {\bibfield  {journal} {\bibinfo  {journal} {Physical Review E}\ }\textbf {\bibinfo {volume} {105}},\ \bibinfo {pages} {014109} (\bibinfo {year} {2022})}\BibitemShut {NoStop}%
\bibitem [{\citenamefont {Ruidas}\ \emph {et~al.}(2026)\citenamefont {Ruidas}, \citenamefont {Roy}, \citenamefont {Bhattacharjee},\ and\ \citenamefont {Moessner}}]{RuidasRoyBhattacharjeeMoessner2026}%
  \BibitemOpen
  \bibfield  {author} {\bibinfo {author} {\bibfnamefont {S.}~\bibnamefont {Ruidas}}, \bibinfo {author} {\bibfnamefont {S.}~\bibnamefont {Roy}}, \bibinfo {author} {\bibfnamefont {S.}~\bibnamefont {Bhattacharjee}},\ and\ \bibinfo {author} {\bibfnamefont {R.}~\bibnamefont {Moessner}},\ }\bibfield  {title} {\bibinfo {title} {Emergence of many-body chaos in the presence of quasiparticles},\ }\href {https://doi.org/10.1103/zkvh-dqqp} {\bibfield  {journal} {\bibinfo  {journal} {Physical Review B}\ }\textbf {\bibinfo {volume} {114}},\ \bibinfo {pages} {014306} (\bibinfo {year} {2026})}\BibitemShut {NoStop}%
\bibitem [{\citenamefont {Bilitewski}\ \emph {et~al.}(2021)\citenamefont {Bilitewski}, \citenamefont {Bhattacharjee},\ and\ \citenamefont {Moessner}}]{BilitewskiBhattacharjeeMoessner2021}%
  \BibitemOpen
  \bibfield  {author} {\bibinfo {author} {\bibfnamefont {T.}~\bibnamefont {Bilitewski}}, \bibinfo {author} {\bibfnamefont {S.}~\bibnamefont {Bhattacharjee}},\ and\ \bibinfo {author} {\bibfnamefont {R.}~\bibnamefont {Moessner}},\ }\bibfield  {title} {\bibinfo {title} {Classical many-body chaos with and without quasiparticles},\ }\href {https://doi.org/10.1103/PhysRevB.103.174302} {\bibfield  {journal} {\bibinfo  {journal} {Physical Review B}\ }\textbf {\bibinfo {volume} {103}},\ \bibinfo {pages} {174302} (\bibinfo {year} {2021})}\BibitemShut {NoStop}%
\bibitem [{\citenamefont {Kim}\ and\ \citenamefont {Huse}(2013)}]{KimHuse2013}%
  \BibitemOpen
  \bibfield  {author} {\bibinfo {author} {\bibfnamefont {H.}~\bibnamefont {Kim}}\ and\ \bibinfo {author} {\bibfnamefont {D.~A.}\ \bibnamefont {Huse}},\ }\bibfield  {title} {\bibinfo {title} {Ballistic spreading of entanglement in a diffusive nonintegrable system},\ }\href {https://doi.org/10.1103/PhysRevLett.111.127205} {\bibfield  {journal} {\bibinfo  {journal} {Physical Review Letters}\ }\textbf {\bibinfo {volume} {111}},\ \bibinfo {pages} {127205} (\bibinfo {year} {2013})}\BibitemShut {NoStop}%
\bibitem [{\citenamefont {Richter}\ and\ \citenamefont {Pal}(2022{\natexlab{b}})}]{PhysRevB.105.L220405}%
  \BibitemOpen
  \bibfield  {author} {\bibinfo {author} {\bibfnamefont {J.}~\bibnamefont {Richter}}\ and\ \bibinfo {author} {\bibfnamefont {A.}~\bibnamefont {Pal}},\ }\bibfield  {title} {\bibinfo {title} {Many-body localization and delocalization dynamics in the thermodynamic limit},\ }\href {https://doi.org/10.1103/PhysRevB.105.L220405} {\bibfield  {journal} {\bibinfo  {journal} {Phys. Rev. B}\ }\textbf {\bibinfo {volume} {105}},\ \bibinfo {pages} {L220405} (\bibinfo {year} {2022}{\natexlab{b}})}\BibitemShut {NoStop}%
\bibitem [{\citenamefont {Pathak}(2026)}]{qpsz-pvxm}%
  \BibitemOpen
  \bibfield  {author} {\bibinfo {author} {\bibfnamefont {T.}~\bibnamefont {Pathak}},\ }\bibfield  {title} {\bibinfo {title} {Exact entanglement dynamics beyond nearest-neighbor dual-unitary floquet systems},\ }\href {https://doi.org/10.1103/qpsz-pvxm} {\bibfield  {journal} {\bibinfo  {journal} {Phys. Rev. B}\ }\textbf {\bibinfo {volume} {114}},\ \bibinfo {pages} {L080306} (\bibinfo {year} {2026})}\BibitemShut {NoStop}%
\bibitem [{\citenamefont {Osipov}\ \emph {et~al.}(2026)\citenamefont {Osipov}, \citenamefont {Spyra}, \citenamefont {Schumann}, \citenamefont {Guhr},\ and\ \citenamefont {Gutkin}}]{osipov2026localcorrelationslongrangedualunitary}%
  \BibitemOpen
  \bibfield  {author} {\bibinfo {author} {\bibfnamefont {V.~A.}\ \bibnamefont {Osipov}}, \bibinfo {author} {\bibfnamefont {M.~C.}\ \bibnamefont {Spyra}}, \bibinfo {author} {\bibfnamefont {J.~C.}\ \bibnamefont {Schumann}}, \bibinfo {author} {\bibfnamefont {T.}~\bibnamefont {Guhr}},\ and\ \bibinfo {author} {\bibfnamefont {B.}~\bibnamefont {Gutkin}},\ }\href {https://arxiv.org/abs/2606.13857} {\bibinfo {title} {Local correlations in long-range dual-unitary kicked hamiltonian chains}} (\bibinfo {year} {2026}),\ \Eprint {https://arxiv.org/abs/2606.13857} {arXiv:2606.13857 [quant-ph]} \BibitemShut {NoStop}%
\bibitem [{\citenamefont {Miyahara}\ and\ \citenamefont {Ueda}(2003)}]{MiyaharaUeda2003}%
  \BibitemOpen
  \bibfield  {author} {\bibinfo {author} {\bibfnamefont {S.}~\bibnamefont {Miyahara}}\ and\ \bibinfo {author} {\bibfnamefont {K.}~\bibnamefont {Ueda}},\ }\bibfield  {title} {\bibinfo {title} {Theory of the orthogonal dimer {Heisenberg} spin model for $\rm {SrCu_{2}(BO_3)_2}$},\ }\href {https://doi.org/10.1088/0953-8984/15/9/201} {\bibfield  {journal} {\bibinfo  {journal} {Journal of Physics: Condensed Matter}\ }\textbf {\bibinfo {volume} {15}},\ \bibinfo {pages} {R327} (\bibinfo {year} {2003})}\BibitemShut {NoStop}%
\bibitem [{\citenamefont {Balents}(2010)}]{Balents2010}%
  \BibitemOpen
  \bibfield  {author} {\bibinfo {author} {\bibfnamefont {L.}~\bibnamefont {Balents}},\ }\bibfield  {title} {\bibinfo {title} {Spin liquids in frustrated magnets},\ }\href {https://doi.org/10.1038/nature08917} {\bibfield  {journal} {\bibinfo  {journal} {Nature}\ }\textbf {\bibinfo {volume} {464}},\ \bibinfo {pages} {199} (\bibinfo {year} {2010})}\BibitemShut {NoStop}%
\bibitem [{\citenamefont {Affleck}(1989)}]{Affleck1989}%
  \BibitemOpen
  \bibfield  {author} {\bibinfo {author} {\bibfnamefont {I.}~\bibnamefont {Affleck}},\ }\bibfield  {title} {\bibinfo {title} {Quantum spin chains and the {H}aldane gap},\ }\href {https://doi.org/10.1088/0953-8984/1/19/001} {\bibfield  {journal} {\bibinfo  {journal} {Journal of Physics: Condensed Matter}\ }\textbf {\bibinfo {volume} {1}},\ \bibinfo {pages} {3047} (\bibinfo {year} {1989})}\BibitemShut {NoStop}%
\bibitem [{\citenamefont {Sen}(2005)}]{Sen2005}%
  \BibitemOpen
  \bibfield  {author} {\bibinfo {author} {\bibfnamefont {D.}~\bibnamefont {Sen}},\ }\bibfield  {title} {\bibinfo {title} {Field theoretic studies of quantum spin systems in one dimension},\ }\href {https://www.iacs.res.in/ijp/ijp_june_05_rev.pdf} {\bibfield  {journal} {\bibinfo  {journal} {Indian Journal of Physics}\ }\textbf {\bibinfo {volume} {79}},\ \bibinfo {pages} {563} (\bibinfo {year} {2005})}\BibitemShut {NoStop}%
\bibitem [{\citenamefont {Budde}\ \emph {et~al.}(2024)\citenamefont {Budde}, \citenamefont {Krstic~Marinkovic},\ and\ \citenamefont {Pinto~Barros}}]{PhysRevD.110.094506}%
  \BibitemOpen
  \bibfield  {author} {\bibinfo {author} {\bibfnamefont {T.}~\bibnamefont {Budde}}, \bibinfo {author} {\bibfnamefont {M.}~\bibnamefont {Krstic~Marinkovic}},\ and\ \bibinfo {author} {\bibfnamefont {J.~C.}\ \bibnamefont {Pinto~Barros}},\ }\bibfield  {title} {\bibinfo {title} {Quantum many-body scars for arbitrary integer spin in $2+1\mathrm{D}$ {A}belian gauge theories},\ }\href {https://doi.org/10.1103/PhysRevD.110.094506} {\bibfield  {journal} {\bibinfo  {journal} {Phys. Rev. D}\ }\textbf {\bibinfo {volume} {110}},\ \bibinfo {pages} {094506} (\bibinfo {year} {2024})}\BibitemShut {NoStop}%
\bibitem [{\citenamefont {Li}\ \emph {et~al.}(2013)\citenamefont {Li}, \citenamefont {Weichselbaum},\ and\ \citenamefont {Delft}}]{PhysRevB.88.245121}%
  \BibitemOpen
  \bibfield  {author} {\bibinfo {author} {\bibfnamefont {W.}~\bibnamefont {Li}}, \bibinfo {author} {\bibfnamefont {A.}~\bibnamefont {Weichselbaum}},\ and\ \bibinfo {author} {\bibfnamefont {J.~V.}\ \bibnamefont {Delft}},\ }\bibfield  {title} {\bibinfo {title} {Identifying symmetry-protected topological order by entanglement entropy},\ }\href {https://doi.org/10.1103/PhysRevB.88.245121} {\bibfield  {journal} {\bibinfo  {journal} {Phys. Rev. B}\ }\textbf {\bibinfo {volume} {88}},\ \bibinfo {pages} {245121} (\bibinfo {year} {2013})}\BibitemShut {NoStop}%
\bibitem [{\citenamefont {Ahmadi}\ \emph {et~al.}(2022)\citenamefont {Ahmadi}, \citenamefont {Abouie}, \citenamefont {Haghshenas},\ and\ \citenamefont {Baeriswyl}}]{PhysRevB.106.174419}%
  \BibitemOpen
  \bibfield  {author} {\bibinfo {author} {\bibfnamefont {N.}~\bibnamefont {Ahmadi}}, \bibinfo {author} {\bibfnamefont {J.}~\bibnamefont {Abouie}}, \bibinfo {author} {\bibfnamefont {R.}~\bibnamefont {Haghshenas}},\ and\ \bibinfo {author} {\bibfnamefont {D.}~\bibnamefont {Baeriswyl}},\ }\bibfield  {title} {\bibinfo {title} {Frustrated mixed-spin ladders: Evidence for a bond-order wave phase between rung-singlet and {H}aldane phases},\ }\href {https://doi.org/10.1103/PhysRevB.106.174419} {\bibfield  {journal} {\bibinfo  {journal} {Phys. Rev. B}\ }\textbf {\bibinfo {volume} {106}},\ \bibinfo {pages} {174419} (\bibinfo {year} {2022})}\BibitemShut {NoStop}%
\bibitem [{\citenamefont {Schliemann}\ and\ \citenamefont {Läuchli}(2012)}]{Schliemann_2012}%
  \BibitemOpen
  \bibfield  {author} {\bibinfo {author} {\bibfnamefont {J.}~\bibnamefont {Schliemann}}\ and\ \bibinfo {author} {\bibfnamefont {A.~M.}\ \bibnamefont {Läuchli}},\ }\bibfield  {title} {\bibinfo {title} {Entanglement spectra of heisenberg ladders of higher spin},\ }\href {https://doi.org/10.1088/1742-5468/2012/11/p11021} {\bibfield  {journal} {\bibinfo  {journal} {Journal of Statistical Mechanics: Theory and Experiment}\ }\textbf {\bibinfo {volume} {2012}},\ \bibinfo {pages} {P11021} (\bibinfo {year} {2012})}\BibitemShut {NoStop}%
\bibitem [{\citenamefont {Pathak}\ and\ \citenamefont {Saha}(2026)}]{pathak2026ergodicitybreakingtransitionqubits}%
  \BibitemOpen
  \bibfield  {author} {\bibinfo {author} {\bibfnamefont {T.}~\bibnamefont {Pathak}}\ and\ \bibinfo {author} {\bibfnamefont {A.~K.}\ \bibnamefont {Saha}},\ }\href {https://arxiv.org/abs/2610.15546} {\bibinfo {title} {Ergodicity breaking transition beyond qubits}} (\bibinfo {year} {2026}),\ \Eprint {https://arxiv.org/abs/2610.15546} {arXiv:2610.15546 [cond-mat.stat-mech]} \BibitemShut {NoStop}%
\bibitem [{\citenamefont {Schulz}(1996)}]{Schulz1996}%
  \BibitemOpen
  \bibfield  {author} {\bibinfo {author} {\bibfnamefont {H.~J.}\ \bibnamefont {Schulz}},\ }\bibfield  {title} {\bibinfo {title} {Dynamics of coupled quantum spin chains},\ }\href {https://doi.org/10.1103/PhysRevLett.77.2790} {\bibfield  {journal} {\bibinfo  {journal} {Physical Review Letters}\ }\textbf {\bibinfo {volume} {77}},\ \bibinfo {pages} {2790} (\bibinfo {year} {1996})}\BibitemShut {NoStop}%
\bibitem [{\citenamefont {Routh}\ \emph {et~al.}(2025)\citenamefont {Routh}, \citenamefont {Ghosh},\ and\ \citenamefont {Kumar}}]{RouthGhoshKumar2025}%
  \BibitemOpen
  \bibfield  {author} {\bibinfo {author} {\bibfnamefont {M.}~\bibnamefont {Routh}}, \bibinfo {author} {\bibfnamefont {S.}~\bibnamefont {Ghosh}},\ and\ \bibinfo {author} {\bibfnamefont {M.}~\bibnamefont {Kumar}},\ }\bibfield  {title} {\bibinfo {title} {Exploring quantum phases in frustrated spin-1/2 chain and ladders: a detailed review},\ }\href {https://doi.org/10.1088/1361-648X/ae0670} {\bibfield  {journal} {\bibinfo  {journal} {Journal of Physics: Condensed Matter}\ }\textbf {\bibinfo {volume} {37}},\ \bibinfo {pages} {403001} (\bibinfo {year} {2025})}\BibitemShut {NoStop}%
\end{thebibliography}%
\end{document}


\raggedbottom
\title{\textcolor{black}{Supplemental Material: Quasi-particle sector hybridization and the onset of chaos in spin ladders}\\
}

\author{Arnav Pushkar\,\,\href{https://orcid.org/0000-0002-1978-8307}
{\includegraphics[scale=0.05]{orcidid.pdf}}}
\email{arnavpushkar@gmail.com}
\affiliation{Department of Physics, Indian Institute of Technology Bombay, Mumbai 400076, India.}
\affiliation{Department of Physics, Kyoto University, Kitashirakawa Oiwakecho, Sakyo-ku, Kyoto 606-8502, Japan}
\author{Abhik Kumar Saha\,\,\href{https://orcid.org/0000-0001-8168-6742}
{\includegraphics[scale=0.05]{orcidid.pdf}}}
\email{saha.abhikkumar.3k@kyoto-u.ac.jp}
\affiliation{Department of Physics, Kyoto University, Kitashirakawa Oiwakecho, Sakyo-ku, Kyoto 606-8502, Japan}

\maketitle
\onecolumngrid
\setcounter{figure}{0}
\renewcommand{\thefigure}{S\arabic{figure}}
\setcounter{figure}{0}
\renewcommand{\theequation}{S\arabic{equation}}
\setcounter{equation}{0}
\renewcommand{\thesection}{S\arabic{section}}
\setcounter{section}{0}

In this supplemental material, we provide details on numerics and additional supporting results. In particular,
\begin{itemize}
    \item In Section \ref{sec:hamiltonian}, we provide the details of the symmetry breaking perturbations included in the Hamiltonian, together with the robustness analysis of the spectral statistics, and discuss the microscopic mechanish of triplon-sector hybridization.
    \item In Section \ref{sec:anisotropy}, we present the dependence of the spectral and triplon diagnostics on the anisotropy parameter $\Delta$. 
    \item In Section \ref{sec:ctt_eev},  we present the eigenstate-resolved connected nearest-neighbor triplon correlations.
    \item In Section \ref{sec:eth_fluctuations}, we show the details of the finite-size scaling of the eigenstate-to-eigenstate fluctuation. 
\end{itemize}

\section{Hamiltonian and Symmetry-Breaking Perturbations}
\label{sec:hamiltonian}

We consider a spin-$1/2$ XXZ ladder containing $N=2\times L$ sites, with periodic boundary conditions along the ladder direction.
The Hamiltonian is given by

\begin{align}
H =
J_{\rm rung} \sum_{i=1}^{L} h^{\rm XXZ}_{(i,1),(i,2)} \nonumber+\sum_{\ell=1}^{2} J_{\rm leg}^{(\ell)} \sum_{i=1}^{L} h^{\rm XXZ}_{(i,\ell),(i+1,\ell)}+H_{\rm diag}+H_{\rm imb},
\label{eq:full_hamiltonian}
\end{align}

where,

\begin{equation}
h^{\rm XXZ}_{a,b}=\frac{1}{2}(S_a^+ S_b^-+S_a^- S_b^+)+\Delta S_a^z S_b^z.
\end{equation}

Here $J_{\rm rung}$ denotes the rung coupling, $J_{\rm leg}$ is the leg coupling, and $\Delta$ is the anisotropy parameter.
The primary tuning parameter is $g=J_{\rm leg}/J_{\rm rung}$. To remove discrete symmetries that would otherwise lead to unresolved symmetry sectors in the spectral analysis, we introduce weak symmetry-breaking perturbations. First, we introduce a small asymmetry between the two leg couplings,
\begin{equation}
J_{\rm leg}^{(1)}=J_{\rm leg}(1+\epsilon_{\rm leg}),\quad
J_{\rm leg}^{(2)}=J_{\rm leg}(1-\epsilon_{\rm leg}),
\label{eq:leg_asymmetry}
\end{equation}
with $\epsilon_{\rm leg}=0.03$. Second, we include a weak diagonal inter-rung coupling along one diagonal orientation,
\begin{equation}
H_{\rm diag}=J_{\rm diag}\sum_{i=1}^{L}h^{\rm XXZ}_{(i,1),(i+1,2)},
\label{eq:diagonal}
\end{equation}
where $J_{\rm diag}=0.07J_{\rm leg}$. Finally, we introduce a uniform leg-antisymmetric magnetic field,
\begin{equation}
H_{\rm imb}=h_{\rm imb}\sum_{i=1}^{L}\left(S^z_{i,1}-S^z_{i,2}\right),
\label{eq:imbalance}
\end{equation}
with $h_{\rm imb}=0.05$. The unequal leg couplings break the symmetry under exchange of the two legs, while the single-orientation diagonal coupling breaks spatial reflection symmetry along the ladder. The leg-antisymmetric field additionally breaks global spin-inversion symmetry.

Moreover in absence of these symmetry-breaking perturbations the triplon-number parity is conserved, meaning that states with even and odd numbers of triplons do not mix. 
The symmetry breaking terms remove this restriction.
In particular, the imbalance field mixes the local rung singlet and $m=0$ triplet states \cite{rungnote}, so that the triplon number is not strictly conserved even in the strong-rung limit. Importantly, these perturbations preserve total $S^z_{\rm tot}$ and translation by one rung. 
Consequently, the numerical calculations are performed in the $S^z_{\rm tot}=0$, $k=0$ symmetry sector, as described in the main text.

\begin{figure}[H]
    \centering
    \includegraphics[width=\linewidth]{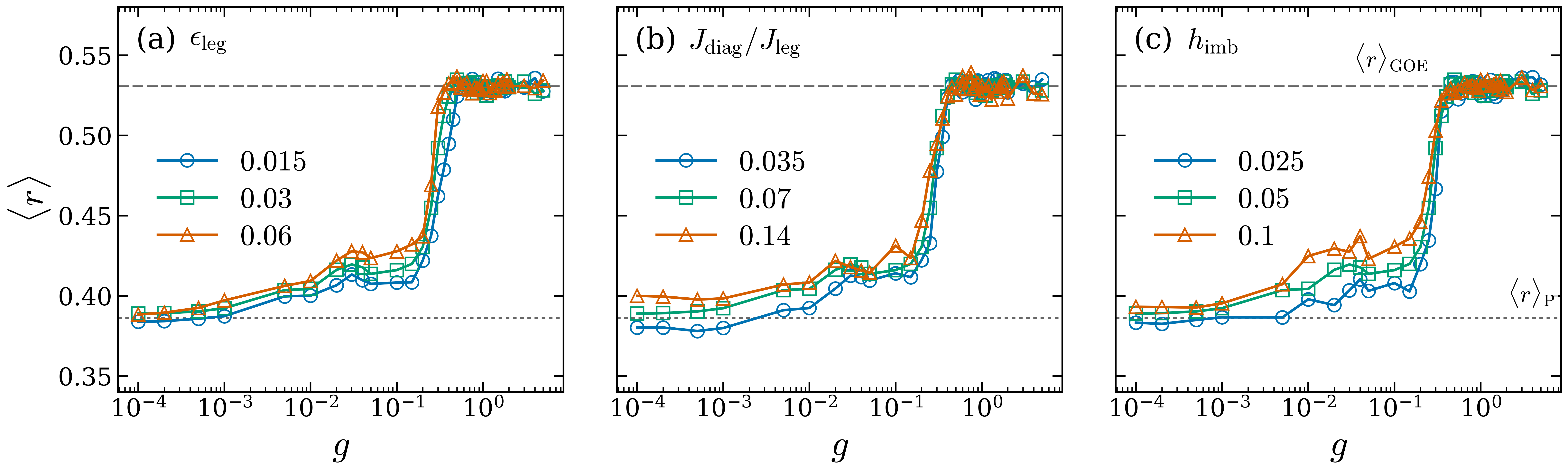}
    \caption{
    Robustness of the average-gap ratio $\langle r\rangle$ against variations of the symmetry-breaking perturbations at $N=2\times10$ and $\Delta=0.8$.
    The panels show $\langle r\rangle$ as a function of $g=J_{\rm leg}/J_{\rm rung}$ upon varying (a) the leg asymmetry $\epsilon_{\rm leg}=0.015,\,0.03,\,0.06$, (b) the relative diagonal coupling $J_{\rm diag}/J_{\rm leg}=0.035,\,0.07,\,0.14$, and (c) the leg-imbalance field $h_{\rm imb}=0.025,\,0.05,\,0.10$.
    In each panel, the other two perturbations are held fixed at their reference values.
    The horizontal dotted and dashed lines denote the Poisson ($\langle r\rangle_{\rm P}\approx0.38$) and GOE ($\langle r\rangle_{\rm GOE}\approx0.5307$) benchmarks, respectively.
    }
    \label{fig:robustness}
\end{figure}
\subsection{Robustness of the Symmetry-Breaking Perturbations}
\label{sec:robustness}
To examine the sensitivity of the spectral crossover to the weak symmetry-breaking perturbations, we consider variations in their strengths while keeping the underlying parameters fixed.
The reference parameters used throughout the main text are $\epsilon_{\rm leg}=0.03$, $J_{\rm diag}/J_{\rm leg}=0.07$, and $h_{\rm imb}=0.05$.
We examine the adjacent average gap ratio $\langle r\rangle$ upon independently varying each perturbation parameter while keeping the other two at their reference values.
The calculations are performed for system size $N=2\times10$ in the $S^z_{\rm tot}=0$, $k=0$ sector using the same central $20\%$ spectral window as described in the main text.

As shown in Fig.~\ref{fig:robustness}, the average gap ratio exhibits a qualitatively similar crossover toward GOE statistics from Poisson statistics for all the considered perturbation strengths.
The curves remain close to one another over a broad range of couplings $g$, with modest quantitative differences near the crossover.
These finite-size results indicate that the emergence of GOE spectral statistics is robust against variations of the symmetry-breaking perturbations and is not restricted to the reference parameter set used in the main text.

\subsection{Rung-parity representation and triplon-sector mixing}

To understand the microscopic processes responsible for triplon-sector hybridization, we introduce the symmetric and antisymmetric spin operators on each rung,
\begin{equation}
\mathbf{T}_i=\mathbf{S}_{i,1}+\mathbf{S}_{i,2},
\qquad
\mathbf{D}_i=\mathbf{S}_{i,1}-\mathbf{S}_{i,2}.
\label{eq:TD_operators}
\end{equation}
Under the exchange of the two spins on a rung, $\mathbf{T}_i$ is even whereas $\mathbf{D}_i$ is odd.
In the local singlet-triplet basis, $\mathbf{T}_i$ acts within the triplet manifold and annihilates the singlet state, while $\mathbf{D}_i$ connects singlet and triplet states.
For example,
\begin{equation}
D_i^z|s\rangle_i=|t_0\rangle_i,
\qquad
D_i^z|t_0\rangle_i=|s\rangle_i.
\label{eq:TD_matrix_elements}
\end{equation}
These properties allow us to distinguish processes that preserve the triplon number from those that couple different triplon-number sectors.

Using $\mathbf{S}_{i,1}=(\mathbf{T}_i+\mathbf{D}_i)/2$ and $\mathbf{S}_{i,2}=(\mathbf{T}_i-\mathbf{D}_i)/2$, the leg interactions can be rewritten as
\begin{equation}
\begin{aligned}
H_{\rm leg}=\frac{J_{\rm leg}}{2}\sum_i\Big[
&\mathbf{T}_i\cdot_\Delta\mathbf{T}_{i+1}
+\mathbf{D}_i\cdot_\Delta\mathbf{D}_{i+1}\\
&+\epsilon_{\rm leg}\big(
\mathbf{T}_i\cdot_\Delta\mathbf{D}_{i+1}
+\mathbf{D}_i\cdot_\Delta\mathbf{T}_{i+1}
\big)
\Big],
\end{aligned}
\label{eq:TD_leg}
\end{equation}
where $\mathbf{A}\cdot_\Delta\mathbf{B}=A^xB^x+A^yB^y+\Delta A^zB^z$ denotes the anisotropic XXZ Hamiltonian term.
The $\mathbf{T}_i\cdot_\Delta\mathbf{T}_{i+1}$ term acts within configurations containing triplets on both neighboring rungs and preserves the total triplon number.
In contrast, the $\mathbf{D}_i\cdot_\Delta\mathbf{D}_{i+1}$ term allows triplon hopping between neighboring rungs as well as pair creation and annihilation, corresponding to changes $\Delta N_t=0,\pm2$.
Consequently, even in the absence of symmetry-breaking perturbations, the leg interactions can hybridize sectors whose triplon numbers differ by two, while preserving triplon-number parity.

The leg asymmetry introduces mixed $\mathbf{T}\mathbf{D}$ terms that connect configurations differing by one triplon, thereby breaking triplon-number parity.
Similarly, the diagonal interaction takes the form
\begin{equation}
\begin{aligned}
H_{\rm diag}=\frac{J_{\rm diag}}{4}\sum_i\Big[
&\mathbf{T}_i\cdot_\Delta\mathbf{T}_{i+1}
-\mathbf{T}_i\cdot_\Delta\mathbf{D}_{i+1}\\
&+\mathbf{D}_i\cdot_\Delta\mathbf{T}_{i+1}
-\mathbf{D}_i\cdot_\Delta\mathbf{D}_{i+1}
\Big],
\end{aligned}
\label{eq:TD_diag}
\end{equation}
which contains both parity-preserving and parity-changing processes.
The leg-antisymmetric field can be written as
\begin{equation}
H_{\rm imb}=h_{\rm imb}\sum_iD_i^z,
\label{eq:TD_imb}
\end{equation}
and directly mixes the local singlet and $m=0$ triplet states, allowing changes in triplon number by one even in the absence of inter-rung coupling.

Thus, the inter-rung interactions provide microscopic channels for the redistribution of triplon occupations, while the symmetry-breaking perturbations additionally couple sectors of opposite triplon-number parity.
In the strong-rung regime, these processes are weak compared with the dominant rung coupling, so the eigenstates can retain an approximate organization according to triplon number.
As the inter-rung coupling increases, the mixing between different rung configurations becomes progressively stronger, leading to the growth of triplon-number fluctuations discussed in the main text.
The operator decomposition therefore provides a microscopic interpretation of the hybridization processes probed by $\Delta N_{t,\alpha}^2$ and the connected triplon correlations.

\section{Dependence on Anisotropy $\Delta$}
\label{sec:anisotropy}

\begin{figure}[H]
\centering
\includegraphics[width=\linewidth]{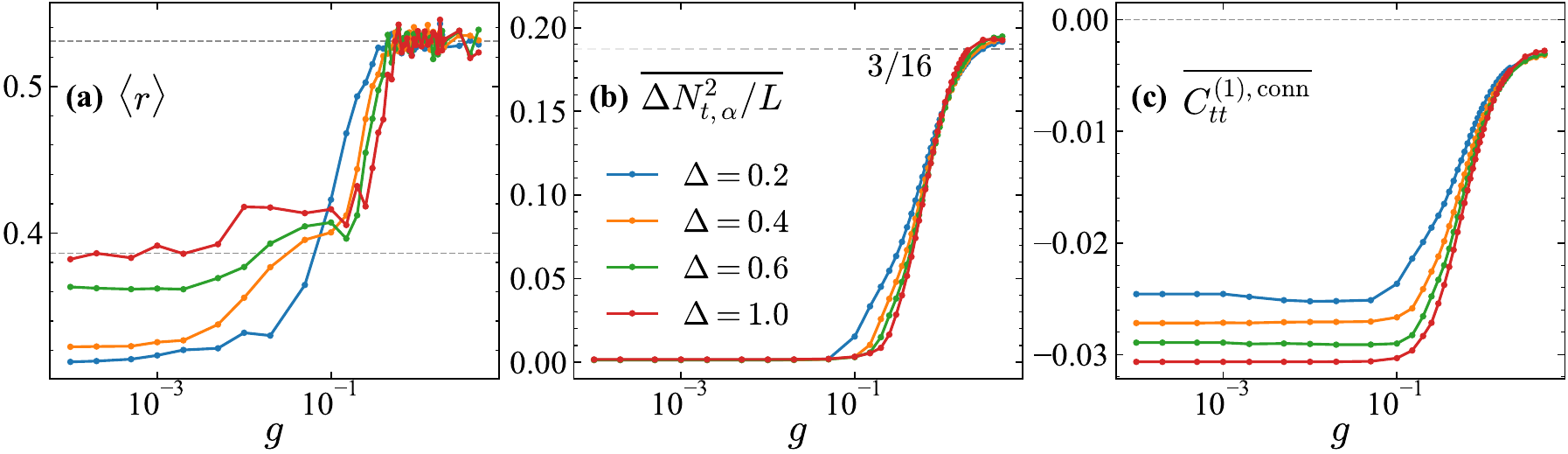}
\caption{
Dependence of the spectral and triplon-sector diagnostics for different values of anisotropy $\Delta$ with system size $N=2\times 9$.
(a) Mean adjacent-gap ratio $\langle r\rangle$, with dashed horizontal lines indicating the Poisson and GOE benchmarks.
(b) Mean triplon-number fluctuation per rung, $\overline{\Delta N_{t, \alpha}^2/L}$, with the thermodynamic infinite-temperature reference value $3/16$ indicated.
(c) Mean connected nearest-neighbor triplon correlation, $\overline{C_{tt}^{(1),\rm conn}}$.
All quantities are evaluated in the $S^z_{\rm tot}=0$, $k=0$ sector over the central spectral window $0.4\leq\mathcal{E}\leq0.6$.
}
\label{fig:supp_anisotropy}
\end{figure}

To examine the dependence of the spectral and triplon-sector diagnostics on the XXZ anisotropy in Fig.~\ref{fig:supp_anisotropy}, we consider $\Delta=0.2,\,0.4,\,0.6,\,1.0$, in addition to $\Delta=0.8$ discussed in the main text. Across all four anisotropies, increasing $g$ drives the system towards the GOE-value, although the small $g$ value for average gap-ratio $\langle r\rangle$ and its crossover value depends on the value of $\Delta$.
The crossover appears to drift towards larger values of $g$ for higher anisotropy parameter $\Delta$.
Similarly the growth of the eigenstate-averaged triplon-number variance per rung $\overline{\Delta N_{t,\alpha}^2}$ with the coupling ratio, indicates hybridization of approximate triplon-number sectors, while the eigenstate-averaged connected nearest-neighbor triplon correlation $\overline{C_{tt}^{(1),\rm conn}}$ approaches zero from negative values. 
The magnitude of $\overline{C_{tt}^{(1),\rm conn}}$ for small values of $g$ decreases as $\Delta$ increases, which is contrary to what we see with respect to system-size $N$ in the main text. 
Notably, the approach toward GOE level statistics precedes the substantial growth of triplon-number fluctuations across the anisotropies considered.
These results suggest that the progressive loss of triplon-sector structure accompanying the emergence of spectral chaos is not restricted to the anisotropy $\Delta=0.8$ discussed in the main text.

\begin{figure}[h]
\centering
\includegraphics[width=0.6\linewidth]{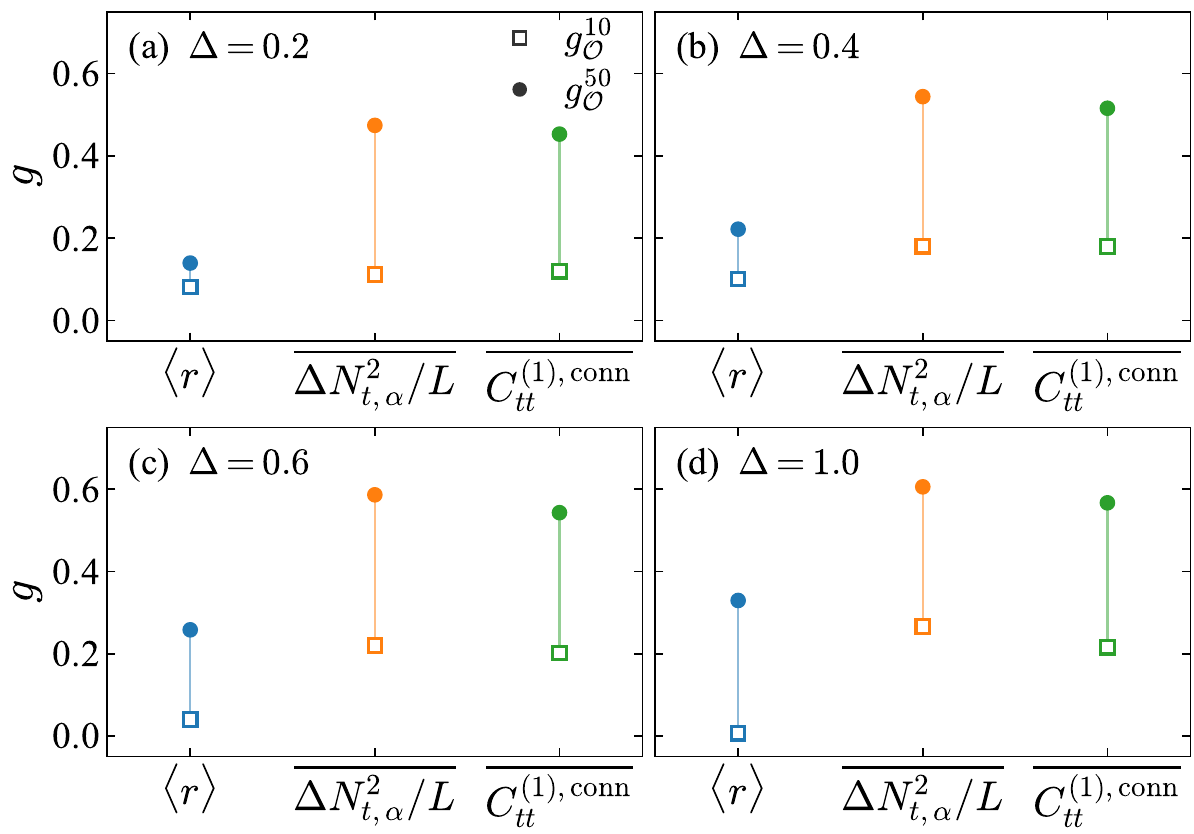}
\caption{
Anisotropy dependence of the crossover scales for the spin-$1/2$ XXZ ladder at $N=2\times 9$.
Panels (a--d) correspond to $\Delta=0.2,0.4,0.6,$ and $1.0$, respectively.
Open squares and filled circles indicate the $10\%$ and $50\%$ crossover scales for the adjacent-gap ratio ($\langle r \rangle$), triplon-number fluctuation $\overline{N_{t,\alpha}^2}$, and connected nearest-neighbor triplon correlation $\overline{C_{tt}^{(1), \rm conn}}$.
The vertical lines connect the two crossover scales for each observable.
All quantities are evaluated in the central spectral window of the $S^z_{\rm tot}=0$, $k=0$ sector.
}
\label{fig:supp_anisotropy_crossovers}
\end{figure}

To quantify the anisotropy dependence of the crossover, Fig.~\ref{fig:supp_anisotropy_crossovers} compares the $10\%$ and $50\%$ crossover scales of the three diagnostics for $N=2\times 9$.
For the average gap ratio, these scales are defined relative to the Poisson and GOE benchmarks, whereas for the triplon observables they are extracted from the changes between their small- and large-$g$ reference values.
Across all four anisotropies, the spectral midpoint remains smaller than the corresponding triplon-sector midpoints, indicating that the spectral crossover develops before substantial triplon-sector reorganization is complete.
Although the characteristic couplings depend on $\Delta$, this separation persists throughout the anisotropy range considered.
At the isotropic point $\Delta=1.0$, the leg-antisymmetric field explicitly breaks the SU(2) symmetry of the exchange Hamiltonian, allowing GOE level statistics to emerge without unresolved total-spin symmetry sectors.

\section{Eigenstate-Resolved Triplon Correlations}
\label{sec:ctt_eev}

Here we present the eigenstate-resolved connected nearest-neighbor triplon correlations $C_{tt}^{(1), \rm conn}$ for representative coupling ratios $g=0.01,\,0.4,\,1.0$, at $\Delta=0.8$.
\begin{figure*}[h]
    \centering
    \includegraphics[width=\textwidth]{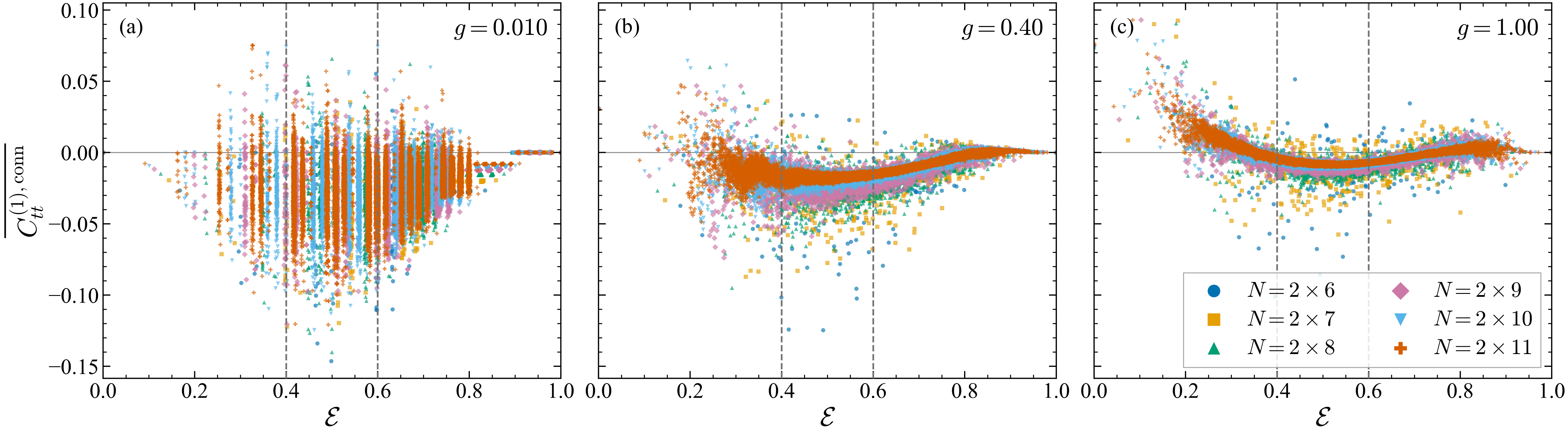}
    \caption{
    Eigenstate-resolved connected nearest-neighbor triplon correlation $C_{tt}^{(1),\mathrm{conn}}$ as a function of normalized energy $\mathcal{E}$ for (a) $g=0.01$, (b) $g=0.4$, and (c) $g=1.0$, at $\Delta=0.8$.
    Different symbols denote system sizes $N=2\times 6$ to $2\times 11$, as indicated in the legend.
    The vertical dashed lines mark the central spectral window $0.4\leq\mathcal{E}\leq0.6$, while the horizontal solid line indicates zero correlation.
    }
    \label{fig:ctt_eev}
\end{figure*}
Fig.~\ref{fig:ctt_eev} shows the evolution of the eigenstate-resolved connected triplon correlations with increasing inter-rung coupling.
In the strong-rung regime ($g=0.01$), the correlations exhibit pronounced eigenstate-to-eigenstate scatter and distinct branch-like structures, reflecting the persistence of approximately conserved rung configurations.
At intermediate coupling ($g=0.4$), a smoother energy-dependent structure emerges, with predominantly negative correlations in the central part of the spectrum and a reduced spread for larger system sizes.
At stronger coupling ($g=1.0$), the eigenstate expectation values become more concentrated around a smooth energy-dependent profile, although appreciable fluctuations remain toward the spectral edges.
These results illustrate the progressive narrowing of eigenstate-to-eigenstate fluctuations with increasing coupling, while the connected triplon correlations retain a nontrivial dependence on energy.
The behavior is consistent with the distinction emphasized in the main text between the onset of GOE level statistics and the evolution of triplon-sector correlations.

\begin{table*}[h]
\centering
\begin{ruledtabular}
\begin{tabular}{ccccccc}
$g$ &
$D_{\mathcal W}^{(16)}$ &
$D_{\mathcal W}^{(18)}$ &
$D_{\mathcal W}^{(20)}$ &
$D_{\mathcal W}^{(22)}$ &
$A$ &
$\gamma$ \\
\hline
0.001 & 519 & 1084 & 4254 & 18564 & $3.339\times10^{-3}$ & 0.1848 \\
0.01  & 481 & 1076 & 4254 & 17240 & $3.212\times10^{-3}$ & 0.1859 \\
0.4   & 655 & 2226 & 7727 & 27053 & $3.279\times10^{-1}$ & 0.4832 \\
1.0   & 782 & 2713 & 9554 & 34071 & $5.998\times10^{-1}$ & 0.5768 \\
1.5   & 775 & 2682 & 9464 & 33896 & $5.864\times10^{-1}$ & 0.5630 \\
\end{tabular}
\end{ruledtabular}
\caption{
Finite-size scaling of the eigenstate-to-eigenstate fluctuations of the triplon-number variance per rung, $\Delta N_{t,\alpha}^2/L$, within the central $20\%$ spectral window at $\Delta=0.8$.
The scaling form $\sigma_O=A D_{\mathcal W}^{-\gamma}$ is fitted using the four largest system sizes, $N=2\times 8,2\times 9,2\times10,2\times 11$.
Here, $D_{\mathcal W}$ denotes the number of eigenstates in the selected window.
}
\label{tab:eth_Dw_alpha}
\end{table*}

\section{ETH Fluctuation Analysis}
\label{sec:eth_fluctuations}

To examine the finite-size scaling of eigenstate-to-eigenstate fluctuations, we consider the triplon-number variance per rung, $O_\alpha=\Delta N_{t,\alpha}^2/L$, within the central $20\%$ spectral window.
The smooth energy dependence of $O_\alpha$ is subtracted to isolate the residual eigenstate fluctuations.
Denoting the smooth background by $f(E_\alpha)$, we define the residuals as
\begin{equation}
\delta O_\alpha=O_\alpha-f(E_\alpha).
\label{eq:supp_residual}
\end{equation}

The standard deviation of these residuals is calculated as
\begin{equation}
\sigma_O=
\sqrt{
\frac{1}{D_{\mathcal W}}
\sum_{\alpha\in\mathcal W}
\left(\delta O_\alpha-\overline{\delta O}\right)^2
},
\label{eq:supp_sigma}
\end{equation}
where $D_{\mathcal W}$ denotes the number of eigenstates within the selected spectral window, and $\overline{\delta O}$ is the mean of the residual background.
We examine the dependence of $\sigma_O$ on $D_{\mathcal W}$ for representative coupling ratios across the spectral crossover.
The reference scaling $\sigma_O\propto D_{\mathcal W}^{-1/2}$~\cite{Srednicki1999,Beugeling_etal_pre_2014} is used to assess the suppression of eigenstate-to-eigenstate fluctuations in the chaotic regime.
To quantify the finite-size trends, we fit the fluctuation amplitude to the power-law form
\begin{equation}
\sigma_O=A D_{\mathcal W}^{-\gamma},
\label{eq:supp_eth_fit}
\end{equation}
where $A$ is the fitted prefactor and $\gamma$ is the scaling exponent.
The fits are performed by linear regression in logarithmic coordinates using the four largest system sizes, $N=2\times 8,2\times 9,2\times 10,2\times 11$, and the resulting parameters are summarized in {Table~\ref{tab:eth_Dw_alpha}}. In the strong-rung regime, $g=0.001$ and $0.01$, we obtain $\gamma\approx0.185$, indicating a substantially slower suppression of eigenstate-to-eigenstate fluctuations than the ETH $D_{\mathcal W}^{-1/2}$ scaling.
At $g=0.4$, the exponent increases to $\gamma\approx0.483$, approaching the expected scaling behavior.
For stronger couplings, $g=1.0$ and $1.5$, the corresponding exponents are $\gamma\approx0.577$ and $0.563$, respectively, remaining close to the reference exponent of $\sim0.50$.

These results demonstrate a pronounced change in the finite-size scaling of triplon-number fluctuations between the strong-rung and larger-coupling regimes.
In particular, the near $D_{\mathcal W}^{-1/2}$ scaling at $g=0.4$ occurs before the triplon-sector diagnostics have reached their large-coupling values, further supporting the separation between the emergence of chaotic spectral behavior and the redistribution of triplon-sector correlations.
Nevertheless, the fitted exponents should be interpreted as finite-size effective exponents rather than definitive asymptotic scaling laws.

\bibliography{refs}